\input harvmac
\def\figin{\epsfcheck\figin}\def\figins{\epsfcheck\figins}
\def\epsfcheck{\ifx\epsfbox\UnDeFiNeD
\message{(NO epsf.tex, FIGURES WILL BE IGNORED)}
\gdef\figin##1{\vskip2in}\gdef\figins##1{\hskip.5in}
\else\message{(FIGURES WILL BE INCLUDED)}%
\gdef\figin##1{##1}\gdef\figins##1{##1}\fi}
\def\DefWarn#1{}
\def\figinsert{\goodbreak\topinsert}
\def\ifig#1#2#3#4{\DefWarn#1\xdef#1{fig.~\the\figno}
\writedef{#1\leftbracket fig.\noexpand~\the\figno}%
\figinsert\figin{\centerline{\epsfxsize=#3mm \epsfbox{#2}}}
\bigskip\medskip\centerline{\vbox{\baselineskip12pt
\advance\hsize by -1truein\noindent\footnotefont{\sl
tFig.~\the\figno:}\sl\ #4}}
\bigskip\endinsert\noindent\global\advance\figno by1}

\def\dbar{{\overline{\partial}}}
\def\Dbar{{\overline{D}}}
\def\ibar{{\bar{\imath}}}

\def\C{{\bf C}}
\def\R{{\bf R}}
\def\Z{{\bf Z}}

\def\a{\alpha}

\def\e{\epsilon}

\def\th{\theta}

\def\l{\lambda}
\def\L{\Lambda}
\def\m{\mu}

\def\t{\tau}

\def\F{\Phi}
\def\w{\omega}

\def\v{\varphi}

\def\P{{\bf P}}

\def\d{\partial}
\def\dbar{{\overline\partial}}
\def\inv{^{-1}}
\def\Tr{{\rm Tr}}
\def\hf{{1\over 2}}
\def\cO{{\cal O}}

\def\cF{{\cal F}}
\def\cN{{\cal N}}
\def\cW{{\cal W}}

\def\cS{{\cal S}}

\def\({\bigl(}
\def\){\bigr)}
\def\<{\langle\,}
\def\>{\,\rangle}

\def\]{\right]}
\def\[{\left[}
\def\vol{{\rm vol}}

\def\tX{{\widetilde{X}}}
\def\tT{{\widetilde{T}}}
\def\tt{{\widetilde{t}}}

\lref\thooft{ G.~'t Hooft, ``A Planar Diagram Theory For Strong
Interactions,'' Nucl.\ Phys.\ B {\bf 72}, 461 (1974).}

\lref\mm{ P.~Ginsparg and G.~W.~Moore, ``Lectures On 2-D Gravity
And 2-D String Theory,'' arXiv:hep-th/9304011.}

\lref\oogvaf{H.~Ooguri and C.~Vafa, ``Two-Dimensional Black Hole
and Singularities of CY Manifolds,''  Nucl.\ Phys.\ B {\bf 463}, 55 (1996).}

\lref\mmm{ P.~Di Francesco, P.~Ginsparg and J.~Zinn-Justin, ``2-D
Gravity and random matrices,'' Phys.\ Rept.\ {\bf 254}, 1 (1995)
[arXiv:hep-th/9306153]}

\lref\kontsevich{M.~Kontsevich, ``Intersection Theory On The
Moduli Space Of Curves And The Matrix Airy Function,'' Commun.\
Math.\ Phys.\ {\bf 147}, 1 (1992).}

\lref\wittentop{E.~Witten, ``On The Structure Of The Topological
Phase Of Two-Dimensional Gravity,'' Nucl.\ Phys.\ B {\bf 340}, 281
(1990)}

\lref\bfss{T.~Banks, W.~Fischler, S.~H.~Shenker and L.~Susskind,
``M theory as a matrix model: A conjecture,'' Phys.\ Rev.\ D {\bf
55}, 5112 (1997) [arXiv:hep-th/9610043].}

\lref\adscft{ O.~Aharony, S.~S.~Gubser, J.~M.~Maldacena, H.~Ooguri
and Y.~Oz, ``Large $N$ field theories, string theory and
gravity,'' Phys.\ Rept.\ {\bf 323}, 183 (2000)
[arXiv:hep-th/9905111]. }

\lref\ghv{D. Ghoshal and C. Vafa, ``$c=1$ string as the
topological theory of the conifold,'' Nucl.\ Phys.\ B {\bf 453},
121 (1995) [arXiv:hep-th/9506122].}

\lref\klmkv{S. Kachru, A. Klemm, W. Lerche, P. Mayr, C. Vafa,
``Nonperturbative Results on the Point Particle Limit of $N=2$
Heterotic String Compactifications,'' Nucl.\ Phys.\ B {\bf 459},
537 (1996) [arXiv:hep-th/9508155].}

\lref\klw{A. Klemm, W. Lerche, P. Mayr, C.Vafa, N. Warner,
``Self-Dual Strings and $N=2$ Supersymmetric Field Theory,''
 Nucl.\ Phys. \ B {\bf 477}, 746 (1996)
[arXiv:hep-th/9604034].}

\lref\kmv{S. Katz, P. Mayr, C. Vafa, ``Mirror symmetry and Exact
Solution of 4D $N=2$ Gauge Theories I,'' Adv.\ Theor.\ Math.\
Phys. {\bf 1}, 53 (1998) [arXiv:hep-th/9706110].}

\lref\gv{R.~Gopakumar and C.~Vafa, ``On the gauge theory/geometry
correspondence,'' Adv.\ Theor.\ Math.\ Phys.\ {\bf 3}, 1415 (1999)
[arXiv:hep-th/9811131]. }

\lref\edel{J.D. Edelstein, K. Oh and R. Tatar, ``Orientifold,
geometric transition and large $N$ duality for $SO/Sp$ gauge
theories,'' JHEP {\bf 0105}, 009 (2001) [arXiv:hep-th/0104037].}

\lref\dasg{K. Dasgupta, K. Oh and R. Tatar, ``Geometric
transition, large $N$ dualities and MQCD dynamics,'' Nucl. Phys.
B {\bf 610}, 331 (2001) [arXiv:hep-th/0105066]\semi -----,
``Open/closed string dualities and Seiberg duality from geometric
transitions in M-theory,'' [arXiv:hep-th/0106040]\semi -----,
``Geometric transition versus cascading solution,'' JHEP {\bf
0201}, 031 (2002) [arXiv:hep-th/0110050].}

\lref\hv{K. Hori and C. Vafa, ``Mirror Symmetry,''
[arXiv:hep-th/0002222].}

\lref\hiv{K. Hori, A. Iqbal and C. Vafa, ``D-Branes And Mirror
Symmetry,'' [arXiv:hep-th/0005247].}

\lref\vaug{C.~Vafa, ``Superstrings and topological strings at
large $N$,'' J.\ Math.\ Phys.\ {\bf 42}, 2798 (2001)
[arXiv:hep-th/0008142].}

\lref\civ{ F.~Cachazo, K.~A.~Intriligator and C.~Vafa, ``A large
$N$ duality via a geometric transition,'' Nucl.\ Phys.\ B {\bf
603}, 3 (2001) [arXiv:hep-th/0103067].}

\lref\ckv{ F.~Cachazo, S.~Katz and C.~Vafa, ``Geometric
transitions and $N = 1$ quiver theories,'' arXiv:hep-th/0108120.}

\lref\cfikv{ F.~Cachazo, B.~Fiol, K.~A.~Intriligator, S.~Katz and
C.~Vafa, ``A geometric unification of dualities,'' Nucl.\ Phys.\ B
{\bf 628}, 3 (2002) [arXiv:hep-th/0110028].}

\lref\cv{ F.~Cachazo and C.~Vafa, ``$N=1$ and $N=2$ geometry from
fluxes,'' arXiv:hep-th/0206017.}

\lref\ov{ H.~Ooguri and C.~Vafa, ``Worldsheet derivation of a
large $N$ duality,'' arXiv:hep-th/0205297.}

\lref\av{ M.~Aganagic and C.~Vafa, ``$G_2$ manifolds, mirror
symmetry, and geometric engineering,'' arXiv:hep-th/0110171.}

\lref\digra{ D.~E.~Diaconescu, B.~Florea and A.~Grassi,
``Geometric transitions and open string instantons,''
arXiv:hep-th/0205234.}

\lref\amv{ M.~Aganagic, M.~Marino and C.~Vafa, ``All loop
topological string amplitudes from Chern-Simons theory,''
arXiv:hep-th/0206164.}

\lref\dfg{ D.~E.~Diaconescu, B.~Florea and A.~Grassi, ``Geometric
transitions, del Pezzo surfaces and open string instantons,''
arXiv:hep-th/0206163.}

\lref\kkl{ S.~Kachru, S.~Katz, A.~E.~Lawrence and J.~McGreevy,
``Open string instantons and superpotentials,'' Phys.\ Rev.\ D
{\bf 62}, 026001 (2000) [arXiv:hep-th/9912151].}

\lref\bcov{ M.~Bershadsky, S.~Cecotti, H.~Ooguri and C.~Vafa,
``Kodaira-Spencer theory of gravity and exact results for quantum
string amplitudes,'' Commun.\ Math.\ Phys.\ {\bf 165}, 311 (1994)
[arXiv:hep-th/9309140]. }

\lref\witcs{ E.~Witten, ``Chern-Simons gauge theory as a string
theory,'' arXiv:hep-th/9207094. }

\lref\naret{I. Antoniadis, E. Gava, K.S. Narain, T.R. Taylor,
``Topological Amplitudes in String Theory,'' Nucl.\ Phys.\ B\ {\bf
413}, 162 (1994) [arXiv:hep-th/9307158].}

\lref\witf{E. Witten, ``Solutions Of Four-Dimensional Field
Theories Via M Theory,'' Nucl.\ Phys.\ B\ {\bf 500}, 3 (1997)
[arXiv:hep-th/9703166].}

\lref\shenker{ S.~H.~Shenker, ``The Strength Of Nonperturbative
Effects In String Theory,'' in Proceedings Cargese 1990, {\it
Random surfaces and quantum gravity}, 191--200. }

\lref\berwa{ M.~Bershadsky, W.~Lerche, D.~Nemeschansky and
N.~P.~Warner, ``Extended $N=2$ superconformal structure of gravity
and W gravity coupled to matter,'' Nucl.\ Phys.\ B {\bf 401}, 304
(1993) [arXiv:hep-th/9211040]}

\lref\akemann{ G.~Akemann, ``Higher genus correlators for the
Hermitian matrix model with multiple cuts,'' Nucl.\ Phys.\ B {\bf
482}, 403 (1996) [arXiv:hep-th/9606004]. }

\lref\wiegmann{P.B. Wiegmann and A. Zabrodin, ``Conformal maps and
integrable hierarchies,'' arXiv:hep-th/9909147.}

\lref\cone{R.~Dijkgraaf and C.~Vafa, to appear.}

\lref\ika{A.~Iqbal and V.~S.~Kaplunovsky, ``Quantum Deconstruction
 of a 5D SYM and its Moduli Space,'' [arXiv:hep-th/0212098].}

\lref\kazakov{ S.~Y.~Alexandrov, V.~A.~Kazakov and I.~K.~Kostov,
``Time-dependent backgrounds of 2D string theory,''
arXiv:hep-th/0205079.}

\lref\dj{ S.~R.~Das and A.~Jevicki, ``String Field Theory And
Physical Interpretation Of $D=1$ Strings,'' Mod.\ Phys.\ Lett.\ A
{\bf 5}, 1639 (1990).}

\lref\givental{ A.B.~Givental, ``Gromov-Witten invariants and
quantization of
 quadratic hamiltonians,'' arXiv:math.AG/0108100.}

\lref\op{ A.~Okounkov and R.~Pandharipande, ``Gromov-Witten
theory, Hurwitz theory, and completed cycle,''
arXiv:math.AG/0204305.}

\lref\dijk{ R.~Dijkgraaf, ``Intersection theory, integrable
hierarchies and topological field theory,'' in Cargese Summer
School on {\it New Symmetry Principles in Quantum Field Theory}
1991, [arXiv:hep-th/9201003].}

\lref\gw{ D.~J.~Gross and E.~Witten, ``Possible Third Order Phase
Transition In The Large $N$ Lattice Gauge Theory,'' Phys.\ Rev.\ D
{\bf 21}, 446 (1980).}

\lref\sw{ N.~Seiberg and E.~Witten, ``Electric-magnetic duality,
monopole condensation, and confinement in $N=2$ supersymmetric
Yang-Mills theory,'' Nucl.\ Phys.\ B {\bf 426}, 19 (1994)
[Erratum-ibid.\ B {\bf 430}, 485 (1994)] [arXiv:hep-th/9407087].}

\lref\kkv{ S.~Katz, A.~Klemm and C.~Vafa, ``Geometric engineering
of quantum field theories,'' Nucl.\ Phys.\ B {\bf 497}, 173 (1997)
[arXiv:hep-th/9609239].}

\lref\taylor{ W.~I.~Taylor, ``D-brane field theory on compact
spaces,'' Phys.\ Lett.\ B {\bf 394}, 283 (1997)
[arXiv:hep-th/9611042].}

\lref\kmmms{ S.~Kharchev, A.~Marshakov, A.~Mironov, A.~Morozov and
S.~Pakuliak, ``Conformal matrix models as an alternative to
conventional multimatrix models,'' Nucl.\ Phys.\ B {\bf 404}, 717
(1993) [arXiv:hep-th/9208044].}

\lref\kostov{ I.~K.~Kostov, ``Gauge invariant matrix model for the
A-D-E closed strings,'' Phys.\ Lett.\ B {\bf 297}, 74 (1992)
[arXiv:hep-th/9208053]. }

\lref\ot{ K.~h.~Oh and R.~Tatar, ``Duality and confinement in
$N=1$ supersymmetric theories from geometric transitions,''
arXiv:hep-th/0112040.}

\lref\ovknot{ H.~Ooguri and C.~Vafa, ``Knot invariants and
topological strings,'' Nucl.\ Phys.\ B {\bf 577}, 419 (2000)
[arXiv:hep-th/9912123]. }

\lref\witk{E.~Witten, ``D-branes And K-Theory,''
JHEP {\bf 9812}, 019 (1998) [arXiv:hep-th/9810188].}

\lref\dvv{ R.~Dijkgraaf, H.~Verlinde and E.~Verlinde, ``Loop
Equations And Virasoro Constraints In Nonperturbative 2-D Quantum
Gravity,'' Nucl.\ Phys.\ B {\bf 348}, 435 (1991).}

\lref\kawai{ M.~Fukuma, H.~Kawai and R.~Nakayama, ``Continuum
Schwinger-Dyson Equations And Universal Structures In
Two-Dimensional Quantum Gravity,'' Int.\ J.\ Mod.\ Phys.\ A {\bf
6}, 1385 (1991).}

\lref\nekrasov{ N.~A.~Nekrasov, ``Seiberg-Witten prepotential from
instanton counting,'' arXiv:hep-th/0206161.}

\lref\ki{ I.~K.~Kostov, ``Bilinear functional equations in 2D
quantum gravity,'' in Razlog 1995, {\it New trends in quantum
field theory}, 77--90, [arXiv:hep-th/9602117].}

\lref\kii{ I.~K.~Kostov, ``Conformal field theory techniques in
random matrix models,'' arXiv:hep-th/9907060.}

\lref\morozov{ A.~Morozov, ``Integrability And Matrix Models,''
Phys.\ Usp.\ {\bf 37}, 1 (1994) [arXiv:hep-th/9303139].}

\lref\fo{ H.~Fuji and Y.~Ookouchi, ``Confining phase
superpotentials for SO/Sp gauge theories via geometric
transition,'' arXiv:hep-th/0205301. }

\lref\marinor{M. Marino, ``Chern-Simons theory, matrix integrals,
and perturbative three-manifold invariants,''
[arXiv:hep-th/0207096].}

\lref\agnt{ I.~Antoniadis, E.~Gava, K.~S.~Narain and T.~R.~Taylor,
``Topological amplitudes in string theory,'' Nucl.\ Phys.\ B {\bf
413}, 162 (1994) [arXiv:hep-th/9307158].}

\lref\vw{ C.~Vafa and E.~Witten, ``A Strong coupling test of S
duality,'' Nucl.\ Phys.\ B {\bf 431}, 3 (1994)
[arXiv:hep-th/9408074].}

\lref\ps{ J.~Polchinski and M.~J.~Strassler, ``The string dual of
a confining four-dimensional gauge theory,''
arXiv:hep-th/0003136.}

\lref\kkn{ V.~A.~Kazakov, I.~K.~Kostov and N.~A.~Nekrasov,
``D-particles, matrix integrals and KP hierarchy,'' Nucl.\ Phys.\
B {\bf 557}, 413 (1999) [arXiv:hep-th/9810035].}

\lref\kpw{ A.~Khavaev, K.~Pilch and N.~P.~Warner, ``New vacua of
gauged $N = 8$ supergravity in five dimensions,'' Phys.\ Lett.\ B
{\bf 487}, 14 (2000) [arXiv:hep-th/9812035].}

\lref\ks{ S.~Kachru and E.~Silverstein, ``4d conformal theories
and strings on orbifolds,'' Phys.\ Rev.\ Lett.\ {\bf 80}, 4855
(1998) [arXiv:hep-th/9802183]. }

\lref\bj{ M.~Bershadsky and A.~Johansen, ``Large $N$ limit of
orbifold field theories,'' Nucl.\ Phys.\ B {\bf 536}, 141 (1998)
[arXiv:hep-th/9803249].}

\lref\lnv{ A.~E.~Lawrence, N.~Nekrasov and C.~Vafa, ``On conformal
field theories in four dimensions,'' Nucl.\ Phys.\ B {\bf 533},
199 (1998) [arXiv:hep-th/9803015]. }

\lref\bkv{ M.~Bershadsky, Z.~Kakushadze and C.~Vafa, ``String
expansion as large $N$ expansion of gauge theories,'' Nucl.\
Phys.\ B {\bf 523}, 59 (1998) [arXiv:hep-th/9803076].}

\lref\vy{ G.~Veneziano and S.~Yankielowicz, ``An Effective
Lagrangian For The Pure $N=1$ Supersymmetric Yang-Mills Theory,''
Phys.\ Lett.\ B {\bf 113}, 231 (1982). }

\lref\sinhv{ S.~Sinha and C.~Vafa, ``$SO$ and $Sp$ Chern-Simons at
large $N$,'' arXiv:hep-th/0012136. }

\lref\aahv{B.~Acharya, M.~Aganagic, K.~Hori and C.~Vafa,
``Orientifolds, mirror symmetry and superpotentials,''
arXiv:hep-th/0202208.}

\lref\dorey{N. Dorey, ``An Elliptic Superpotential for Softly
Broken $N=4$ Supersymmetric Yang-Mills,'' JHEP {\bf 9907}, 021
(1999) [arXiv:hep-th/9906011].}

\lref\ls{ R.~G.~Leigh and M.~J.~Strassler, ``Exactly marginal
operators and duality in four-dimensional $N=1$ supersymmetric
gauge theory,'' Nucl.\ Phys.\ B {\bf 447}, 95 (1995)
[arXiv:hep-th/9503121]. }

\lref\klyt{ A.~Klemm, W.~Lerche, S.~Yankielowicz and S.~Theisen,
``Simple singularities and $N=2$ supersymmetric Yang-Mills
theory,'' Phys.\ Lett.\ B {\bf 344}, 169 (1995)
[arXiv:hep-th/9411048]}

\lref\af{ P.~C.~Argyres and A.~E.~Faraggi, ``The vacuum structure
and spectrum of $N=2$ supersymmetric $SU(n)$ gauge theory,''
Phys.\ Rev.\ Lett.\ {\bf 74}, 3931 (1995) [arXiv:hep-th/9411057].}

\lref\mo{C.~Montonen and D.~I.~Olive, ``Magnetic Monopoles As
Gauge Particles?,'' Phys.\ Lett.\ B {\bf 72}, 117 (1977). }

\lref\gvw{S.~Gukov, C.~Vafa and E.~Witten, ``CFT's from Calabi-Yau
four-folds,'' Nucl.\ Phys.\ B {\bf 584}, 69 (2000) [Erratum-ibid.\
B {\bf 608}, 477 (2001)] [arXiv:hep-th/9906070]. }

\lref\tv{ T.~R.~Taylor and C.~Vafa, ``RR flux on Calabi-Yau and
partial supersymmetry breaking,'' Phys.\ Lett.\ B {\bf 474}, 130
(2000) [arXiv:hep-th/9912152]. }

\lref\mayr{ P.~Mayr, ``On supersymmetry breaking in string theory
and its realization in brane worlds,'' Nucl.\ Phys.\ B {\bf 593},
99 (2001) [arXiv:hep-th/0003198]. }

\lref\qfvy{A.~de la Macorra and G.G.~Ross, Nucl.\ Phys.\ B {\bf
404}, 321 (1993)\semi C. P. Burgess, J.-P. Derendinger, F.
Quevedo, M. Quiros, ``On Gaugino Condensation with Field-Dependent
Gauge Couplings,'' Annals Phys. {\bf 250}, 193 (1996)
[arXiv:hep-th/9505171].}

\lref\div{ A. D'Adda, A.C. Davis, P. Di Vecchia and P. Salomonson,
Nucl. Phys. B {\bf 222}, 45 (1983).}

\lref\berva{ N. Berkovits and C. Vafa, ``$N=4$ Topological
Strings,'' Nucl. Phys. B {\bf 433} 123 (1995)
[arXiv:hep-th/9407190].}

\lref\novikov{ V.~A.~Novikov, M.~A.~Shifman, A.~I.~Vainshtein and
V.~I.~Zakharov, ``Instanton Effects In Supersymmetric Theories,''
Nucl.\ Phys.\ B {\bf 229}, 407 (1983). }

\lref\dk{ N.~Dorey and S.~P.~Kumar, ``Softly-broken $N = 4$
supersymmetry in the large-$N$ limit,'' JHEP {\bf 0002}, 006
(2000) [arXiv:hep-th/0001103]. }

\lref\ds{ N.~Dorey and A.~Sinkovics, ``$N = 1^*$ vacua, fuzzy
spheres and integrable systems,'' JHEP {\bf 0207}, 032 (2002)
[arXiv:hep-th/0205151]. }

\lref\ads{ I.~Affleck, M.~Dine and N.~Seiberg, ``Supersymmetry
Breaking By Instantons,'' Phys.\ Rev.\ Lett.\ {\bf 51}, 1026
(1983). }

\lref\bele{ D. Berenstein, V. Jejjala and R. G. Leigh, ``Marginal
and Relevant Deformations of $N=4$ Field Theories and
Non-Commutative Moduli Spaces of Vacua,'' Nucl.Phys. B {\bf 589}
196 (2000) [arXiv:hep-th/0005087].}

\lref\gins{ P. Ginsparg, ``Matrix models of 2d gravity,'' Trieste
Lectures (July, 1991), Gava et al., 1991 summer school in H.E.P.
and Cosmo. [arXiv:hep-th/9112013].}

\lref\kostovsix{ I. Kostov, ``Exact Solution of the Six-Vertex
Model on a Random Lattice,'' Nucl.Phys. B {\bf 575} 513 (2000)
 [arXiv:hep-th/9911023].}

\lref\hoppe{ J.~Goldstone, unpublished; J.~Hoppe, ``Quantum theory
of a massless relativistic surface,'' MIT PhD thesis, 1982.}

\lref\dvi{R.~Dijkgraaf, C.~Vafa, ``Matrix Models, Topological
Strings, and Supersymmetric Gauge Theories,'' Nucl. Phys. {\bf
B644} (2002) 3--20, [arXiv:hep-th/0206255].}

\lref\dvii{R.~Dijkgraaf, C.~Vafa, ``On Geometry and Matrix
Models,'' Nucl. Phys. {\bf B644} (2002) 21--39,
[arXiv:hep-th/0207106].}

\lref\dviii{R.~Dijkgraaf, C.~Vafa, ``A Perturbative Window into
Non-Perturbative Physics,'' arXiv:hep-th/0208048.}

\lref\dglvz{ R.~Dijkgraaf, M.T.~Grisaru, C.S.~Lam, C.~Vafa, and
D.~Zanon, ``Perturbative Computation of Glueball
Superpotentials,'' [arXiv:hep-th/0211017]. }

\lref\cdsw{
F.~Cachazo, M.~R.~Douglas, N.~Seiberg and E.~Witten, ``Chiral rings
and anomalies in supersymmetric gauge theory,'' JHEP {\bf 0212}, 071
(2002) [arXiv:hep-th/0211170].}

\lref\dgkv{ R.~Dijkgraaf, S.~Gukov, V.A.~Kazakov, C.~Vafa,
``Perturbative Analysis of Gauged Matrix Models,''
[arXiv:hep-th/0210238].}

\lref\wittft{ E.~Witten, ``Topological quantum field theory,''
Commun. Math. Phys. {\bf 117} (1988) 353.}

\lref\witsdual{ E.~Witten, ``On S duality in Abelian gauge
theory,'' arXiv:hep-th/9505186}

\lref\mw{ G.~W.~Moore and E.~Witten, ``Integration over the
u-plane in Donaldson theory,'' Adv.\ Theor.\ Math.\ Phys.\ {\bf
1}, 298 (1998) [arXiv:hep-th/9709193].}

\lref\mamo{ M.~Marino and G.~W.~Moore, ``Integrating over the
Coulomb branch in $N = 2$ gauge theory,'' Nucl.\ Phys.\ Proc.\
Suppl.\ {\bf 68}, 336 (1998) [arXiv:hep-th/9712062]; ``The
Donaldson-Witten function for gauge groups of rank larger than
one,'' Commun.\ Math.\ Phys.\ {\bf 199}, 25 (1998)
[arXiv:hep-th/9802185]; ``Donaldson invariants for non-simply
connected manifolds,'' [arXiv:hep-th/9804104].}

\lref\star{ G.~W.~Moore, ``Matrix Models Of 2-D Gravity And
Isomonodromic Deformation,'' lectures at 1990 Cargese Workshop on
{\sl Random Surfaces, Quantum Gravity and Strings,} Prog.\ Theor.\
Phys.\ Suppl.\ {\bf 102}, 255 (1990). }

\lref\zam{ A.~B.~Zamolodchikov, ``Conformal Scalar Field On The
Hyperelliptic Curve And Critical Ashkin-Teller Multipoint
Correlation Functions,'' Nucl.\ Phys.\ B {\bf 285}, 481 (1987). }

\lref\dfms{ L.~J.~Dixon, D.~Friedan, E.~J.~Martinec and
S.~H.~Shenker, ``The Conformal Field Theory Of Orbifolds,'' Nucl.\
Phys.\ B {\bf 282}, 13 (1987). }

\lref\br{ M.~A.~Bershadsky and A.~O.~Radul, ``Conformal Field
Theories With Additional $Z_N$ Symmetry,'' Sov.\ J.\ Nucl.\
Phys.\ {\bf 47}, 363 (1988) [Yad.\ Fiz.\ {\bf 47}, 575 (1988)]. }

\lref\lns{ A.~Losev, N.~Nekrasov and S.~L.~Shatashvili, ``Issues
in topological gauge theory,'' Nucl.\ Phys.\ B {\bf 534}, 549
(1998) [arXiv:hep-th/9711108].}

\lref\ackm{ J.~Ambjorn, L.~Chekhov, C.~F.~Kristjansen and
Y.~Makeenko, ``Matrix model calculations beyond the spherical
limit,'' Nucl.\ Phys.\ B {\bf 404}, 127 (1993) [Erratum-ibid.\ B
{\bf 449}, 681 (1995)] [arXiv:hep-th/9302014].}

\lref\seiberg{
N.~Seiberg, ``Electric-magnetic duality in supersymmetric nonAbelian
gauge theories,'' Nucl.\ Phys.\ B {\bf 435}, 129 (1995)
[arXiv:hep-th/9411149].}

\lref\mss{
N.~Marcus, A.~Sagnotti and W.~Siegel, ``Ten-Dimensional Supersymmetric
Yang-Mills Theory In Terms Of Four-Dimensional Superfields,'' Nucl.\
Phys.\ B {\bf 224}, 159 (1983).}

\lref\agw{
N.~Arkani-Hamed, T.~Gregoire and J.~Wacker, ``Higher dimensional
supersymmetry in 4D superspace,'' JHEP {\bf 0203}, 055 (2002)
[arXiv:hep-th/0101233].}

\lref\vafadd{
C.~Vafa, ``Brane/anti-brane systems and U(N$|$M) supergroup,''
arXiv:hep-th/0101218.  }

\lref\muet{
D.~Ghoshal, D.~P.~Jatkar and S.~Mukhi, ``Kleinian singularities and
the ground ring of $c=1$ string theory,'' Nucl.\ Phys.\ B {\bf 395},
144 (1993) [arXiv:hep-th/9206080].  }

\lref\ura{ A.~M.~Uranga, ``Brane
Configurations for Branes at Conifolds,'' JHEP {\bf 9901}, 022 (1999)
[arXiv:hep-th/9811004].}

\lref\gn{
S.~Gubser, N.~Nekrasov and S.~Shatashvili, ``Generalized conifolds and
four dimensional N = 1 superconformal theories,'' JHEP {\bf 9905}, 003
(1999) [arXiv:hep-th/9811230].}

\lref\akmv{
M.~Aganagic, A.~Klemm, M.~Marino and C.~Vafa, ``Matrix model as a
mirror of Chern-Simons theory,'' arXiv:hep-th/0211098.}

\lref\wadia{
S.~R.~Wadia, ``$N =\infty$ Phase Transition In A Class Of Exactly
Soluble Model Lattice Gauge Theories,'' Phys.\ Lett.\ B {\bf 93}, 403
(1980).  }

\lref\gk{
D.~J.~Gross and I.~R.~Klebanov, ``One-Dimensional String Theory On A
Circle,'' Nucl.\ Phys.\ B {\bf 344}, 475 (1990).}

\lref\witYM{E.~Witten,
``Two-dimensional gauge theories revisited,'' J.\ Geom.\ Phys.\ {\bf
9}, 303 (1992) [arXiv:hep-th/9204083].}

\lref\gt{
D.~J.~Gross and W.~I.~Taylor, ``Two-dimensional QCD is a string
theory,'' Nucl.\ Phys.\ B {\bf 400}, 181 (1993)
[arXiv:hep-th/9301068].}

\lref\nareatal{
I.~Antoniadis, E.~Gava, K.~S.~Narain and T.~R.~Taylor, ``$N=2$ type II
heterotic duality and higher derivative F terms,'' Nucl.\ Phys.\ B
{\bf 455}, 109 (1995) [arXiv:hep-th/9507115].  }

\lref\normal{
I.~K.~Kostov, I.~Krichever, M.~Mineev-Weinstein, P.~B.~Wiegmann and
A.~Zabrodin, ``$\tau$-function for analytic curves,''
arXiv:hep-th/0005259\semi V.~A.~Kazakov and A.~Marshakov, ``Complex
curve of the two matrix model and its tau-function,''
arXiv:hep-th/0211236.  }

\lref\dm{
M.~R.~Douglas and G.~W.~Moore, ``D-branes, Quivers, and ALE
Instantons,'' arXiv:hep-th/9603167.}

\lref\supermatrix{
S.~A.~Yost, ``Supermatrix models,'' Int.\ J.\ Mod.\ Phys.\ A {\bf 7},
6105 (1992) [arXiv:hep-th/9111033].}

\lref\war{M. Bershadsky, W.~Lerche, D.~Nemeschansky, N.P.~Warner,
``Extended N=2 Superconformal Structure of Gravity and
W-Gravity Coupled to Matter,'' Nucl.\ Phys. \ B {\bf 401}, 304 (1993)
 [arXiv:hep-th/9211040].}

\lref\witz{
E.~Witten and B.~Zwiebach, ``Algebraic structures and differential
geometry in $2-D$ string theory,'' Nucl.\ Phys.\ B {\bf 377}, 55
(1992) [arXiv:hep-th/9201056].}

\lref\klebs{
I.~R.~Klebanov and M.~J.~Strassler, ``Supergravity and a confining
gauge theory: Duality cascades and chiSB-resolution of naked
singularities,'' JHEP {\bf 0008}, 052 (2000) [arXiv:hep-th/0007191].}

\lref\kw{
I.~R.~Klebanov and E.~Witten, ``Superconformal field theory on
threebranes at a Calabi-Yau singularity,'' Nucl.\ Phys.\ B {\bf 536},
199 (1998) [arXiv:hep-th/9807080]\semi -----, ``AdS/CFT correspondence
and symmetry breaking,'' Nucl.\ Phys.\ B {\bf 556}, 89 (1999)
[arXiv:hep-th/9905104].}

\lref\nima{
N.~Arkani-Hamed, A.~G.~Cohen, D.~B.~Kaplan, A.~Karch and L.~Motl,
``Deconstructing (2,0) and little string theories,''
arXiv:hep-th/0110146.  }

\lref\seibergL{
N.~Seiberg, ``Notes On Quantum Liouville Theory And Quantum Gravity,''
Prog.\ Theor.\ Phys.\ Suppl.\ {\bf 102}, 319 (1990).  }

\lref\ceresole{
A.~Ceresole, G.~Dall'Agata, R.~D'Auria and S.~Ferrara, ``Spectrum of
type IIB supergravity on $AdS_5\times T^{1,1}$: Predictions on $N = 1$
SCFT's,'' Phys.\ Rev.\ D {\bf 61}, 066001 (2000)
[arXiv:hep-th/9905226].  }

\lref\rand{
D.~P.~Jatkar and S.~Randjbar-Daemi, ``Type IIB string theory on $AdS_5
\times T^{n,n'}$,'' Phys.\ Lett.\ B {\bf 460}, 281 (1999)
[arXiv:hep-th/9904187].  }

\lref\witg{
E.~Witten, ``Ground ring of two-dimensional string theory,'' Nucl.\
Phys.\ B {\bf 373}, 187 (1992) [arXiv:hep-th/9108004].}

\lref\kp{
I.~R.~Klebanov and A.~M.~Polyakov, ``Interaction of discrete states in
two-dimensional string theory,'' Mod.\ Phys.\ Lett.\ A {\bf 6}, 3273
(1991) [arXiv:hep-th/9109032].}

\lref\kg{
S.~S.~Gubser and I.~R.~Klebanov, ``Baryons and domain walls in an $N =
1$ superconformal gauge theory,'' Phys.\ Rev.\ D {\bf 58}, 125025
(1998) [arXiv:hep-th/9808075].  }

\lref\kt{
I.~R.~Klebanov and A.~A.~Tseytlin, ``Gravity duals of supersymmetric
$SU(N) \times SU(N+M)$ gauge theories,'' Nucl.\ Phys.\ B {\bf 578},
123 (2000) [arXiv:hep-th/0002159].}

\lref\mv{
S.~Mukhi and C.~Vafa, ``Two-dimensional black hole as a topological
coset model of $c = 1$ string theory,'' Nucl.\ Phys.\ B {\bf 407}, 667
(1993) [arXiv:hep-th/9301083].  }

\lref\minsei{
S.~Minwalla and N.~Seiberg, ``Comments on the IIA NS5-brane,'' JHEP
{\bf 9906}, 007 (1999) [arXiv:hep-th/9904142].}

\lref\ims{
K.~A.~Intriligator, D.~R.~Morrison and N.~Seiberg, ``Five-dimensional
supersymmetric gauge theories and degenerations of Calabi-Yau
spaces,'' Nucl.\ Phys.\ B {\bf 497}, 56 (1997)
[arXiv:hep-th/9702198].}.

\lref\acg{
N.~Arkani-Hamed, A.~G.~Cohen and H.~Georgi, ``(De)constructing
dimensions,'' Phys.\ Rev.\ Lett.\ {\bf 86}, 4757 (2001)
[arXiv:hep-th/0104005].}

\Title
{\vbox{
\hbox{hep-th/0302011}
\hbox{CALT-68-2425}
\hbox{HUTP-03/A010}
\hbox{ITFA-2003-7}
\vskip-20mm
}}
{\vbox{
\centerline{${\cal N}=1$ Supersymmetry, Deconstruction}
\vskip4mm
\centerline{ and Bosonic Gauge Theories}
}}
\vskip-6mm
\centerline{Robbert Dijkgraaf}
\vskip.05in
\centerline{\sl Institute for Theoretical Physics \&}
\centerline{\sl Korteweg-de Vries Institute for Mathematics}
\centerline{\sl University of Amsterdam, Amsterdam, The Netherlands}
\smallskip
\centerline{and}
\smallskip
\centerline{Cumrun Vafa} \vskip.05in \centerline{\sl Dept. of Physics,
Caltech, Pasadena, CA 91125, USA}
 \centerline{\sl and}
\centerline{\sl Dept. of Physics, Harvard University, Cambridge, MA 02138, USA}
\smallskip

\centerline{\bf Abstract}

We show how the full holomorphic geometry of local Calabi-Yau
threefold compactifications with ${\cal N}=1$ supersymmetry can be
obtained from matrix models.  In particular for the conifold geometry
we relate F-terms to the general amplitudes of $c=1$ non-critical
bosonic string theory, and express them in a quiver or, equivalently,
super matrix model. Moreover we relate, by deconstruction, the
uncompactified $c=1$ theory to the six-dimensional conformal $(2,0)$
theory.  Furthermore, we show how we can use the idea of
deconstruction to connect $4+k$ dimensional supersymmetric gauge
theories to a $k$-dimensional internal bosonic gauge theory,
generalizing the relation between 4d theories and matrix
models. Examples of such bosonic systems include unitary matrix models
and gauged matrix quantum mechanics, which deconstruct 5-dimensional
supersymmetric gauge theories, and Chern-Simons gauge theories, which
deconstruct gauge theories living on branes wrapped over cycles in
Calabi-Yau threefolds.


\Date{February 2003}

\newsec{Introduction}

In a series of papers \refs{\dvi,\dvii,\dviii} we have advanced a
connections between a large class of ${\cal N}=1$ gauge theories in
four dimensions and matrix models. This connection was motivated from
insights coming from string theory \refs{\bcov ,\gv ,\vaug ,\civ
,\ov}.  Moreover, inspired by the string theory derivation of this
result a direct field theory proof has been obtained from surprisingly
simple computations \dglvz, where one can see how diagram by diagram
the field theory computation reduces to the combinatorics of planar
diagrams of the corresponding matrix model. An alternative derivation
has been given in \cdsw\ based on a generalization of the Konishi
anomaly.

However, our original interest, that led to the works
\refs{\dvi,\dvii,\dviii}, was to find a direct connection between
matrix models and non-critical bosonic strings on the one hand, and
topological strings and $\cN=1$ supersymmetric gauge theories on the
other.  Since this motivation may not be familiar to many readers we
will briefly review it now.

Non-critical bosonic strings were heavily investigated more than ten
years ago with some very striking results. `Non-critical' refers here
to the fact that the dimension of the target space is not 26, but less
than that. It was found that for dimension (central charge) less than
or equal to 1, one can compute string amplitudes exactly using double
scaling limits of matrix models. The limiting case of $c=1$ was
precisely the limit where the usual bosonic `tachyon field' was still
non-tachyonic --- in fact massless. Many papers were devoted to
studying non-critical bosonic strings with $c=1$, the target space
represented by a circle of radius $R$ (including the
decompactification limit $R\rightarrow \infty$). For a review see
\mm.

An interesting connection between non-critical strings and topological
strings \wittentop\ in the context of twisted $\cN=2$ SCFT's was
found in \war\ where using the string BRST complex, the $c=1$ theories
were mapped to $\cN=2$ theories with ${\widehat c}=3$, {\it
i.e.}\ theories with the same central charge as supersymmetric sigma
models on Calabi-Yau 3-folds. Further evidence for this correspondence
was found in \bcov\ where it was seen that the scaling properties of
topological string amplitudes near a deformed conifold singularity are
exactly the same as that for $c=1$ non-critical strings.

Properties of non-critical bosonic string on a circle strongly depends
on the radius of the circle. In particular it was shown in
\refs{\witg,\kp} that the string theory enjoys an infinite enhanced symmetry
algebra in the target space at the self-dual radius. Moreover, it was
shown that a real three-dimensional geometry captures the ring of
observables of this theory \refs{\witg,\witz}. This three-dimensional
geometry is a real version of the deformed conifold.

The connection between $c=1$ non-critical bosonic strings and ${\cal
N}=2$ superconformal field theories was made more concrete in \mv\
where it was shown that $c=1$ on a circle of self-dual radius is given
by an $\cN=2$ Kazama-Suzuki GKO coset model $SL(2)/U(1)$ at level
$3$. Later, it was shown in \ghv\ that this corresponds to topological
B-model on the Calabi-Yau threefold given by the deformed
conifold. The chiral ring of the non-critical string in this context
gets mapped to the ring of holomorphic functions on the manifold,
which is the chiral ring of topological B-model. Note that it is
crucial that observables of the B-model involve only holomorphic
functions and so this can be mapped to the ring of observables of {\it
bosonic} strings at self-dual radius.

On the other hand, in trying to embed the large $N$ topological string
duality of \gv\ in superstrings one ends up with the result that the
topological B-model of the conifold describes IR properties of a pure
$\cN=1$ supersymmetric Yang-Mills \vaug. Thus it was conjectured in
\vaug\ that the totality of F-terms of the $\cN=1$ theory of pure
Yang-Mills is captured by non-critical bosonic strings with $c=1$ at
self-dual radius.

On the other hand, following a completely different path it was also
found in \klebs\ that the deformed conifold is
equivalent in some limit to the IR properties of pure $\cN=1 $
Yang-Mills. This was obtained by considering a non-conformal
deformation of the $\cN=1$ AdS/CFT duality found in
\kw, initiated in \refs{\kg,\kt}.
However the pure $\cN=1$ emerged only at the end of a cascade of
Seiberg-like dualities. In particular if one goes to higher energies,
one finds a different gauge theory description with more degrees of
freedom giving rise to an affine ${\widehat A}_1$ quiver theory. The
duality cascade was interpreted in \cfikv\ as affine Weyl reflections
of ${\widehat A}_1$ (corresponding to generalized flops).

One aim of this note is to realize the suggestion of \vaug, relating
$\cN=1$ SYM and $c=1$ strings, in terms of the $\cN=1$ affine
${\widehat A}_1$ quiver theory. In other words, we argue that the
F-terms of the affine ${\widehat A}_1$ theory are equivalent to
correlation functions of the non-critical bosonic string at the
self-dual radius. Moreover, using the recent results
\refs{\dvi,\dvii,\dviii} we can then show that this in turn is
captured by a (quiver) matrix model. This leads to a new realization
of $c=1$ theory at self-dual radius in terms of a matrix model (not in
the double scaling sense, but \`a la 't Hooft). We also show that this
model can be interpreted as a {\it super}matrix model that naturally
appears from a topological brane/antibrane system. Moreover we extend
this dictionary by showing that the non-critical $c=1$ bosonic string
at $k$ times the self-dual radius captures the F-terms of an
${\widehat A_{2k-1}}$ affine quiver theory. In particular $c=1$
bosonic string on a non-compact line is related to affine
$\widehat{A}_\infty$ quiver theory. This is indeed interesting as this
latter theory in turn is related to the $(2,0)$ little strings by
deconstruction in a certain range of parameters \nima.

Another aim of this paper is to reconstruct the full Calabi-Yau
threefold geometry from gauge theory. In particular with the affine
quiver theories all the holomorphic correlations of the $\cN=1$
theory can be computed from matrix models. Given the equivalence of
these with string theories, we can thus directly compute all the
relevant holomorphic quantities of the string theory directly, and
easily, using the matrix model. Moreover, one may naturally expect
that this captures (at the conformal point) even the structure of the
D-term and so it implicitly characterizes the full string theory. This
is somewhat analogous to the two-dimensional program that F-term data
in $\cN=2$ theories characterizes the SCFT completely (with a
unique compatible D-term). We thus come up with a potential dramatical
reformulation of the full string theory in terms of simple matrix
models, in an implicit way.

\bigskip

We also investigate the generalization of matrix models to higher
dimensional gauge theories and their meaning in the context of $\cN=1$
supersymmetric gauge theories in 4 dimensions.  This
is done by viewing the internal theory as an infinite collections of
4d multiplets coupled in a complicated, but computable, way.  More
generally, we show that any bosonic gauge theory in $k$ dimensions
gives rise to some $\cN=1$ dynamics in 4d, perhaps with infinitely
many massive multiplets, interacting in a complicated way.  This turns
out to be a powerful way to deconstruct higher-dimensional theories.

In particular we map matrix quantum mechanics to the deconstruction of
$\cN=1$ supersymmetric Yang-Mills in 5 dimensions interacting with
matter hypermultiplets.  Furthermore we map Chern-Simons gauge theory
in the real or holomorphic version, to the dynamics of gauge theory on
D6 branes wrapped over 3-cycles or D9 brane wrapped over Calabi-Yau
3-fold respectively, directly from a 4-dimensional point of view.
However the aim of using deconstruction in our context is somewhat
different from that used in \acg. In particular our aim in relating
the higher-dimensional theory to a 4-dimensional theory is not to
provide a potential UV completion of the higher-dimensional theory,
but rather reduce its F-term content to 4 dimensions where F-term
computations are simple as exemplified by the connection between
matrix models and supersymmetric gauge theories.

\bigskip

The organization of this paper is as follows: In section 2 we review
some basic aspects of $c=1$ non-critical strings. In section 3 we
explain our proposal of how this is related to a matrix model. In
section 4 we show how to identify the relevant $\cN=1$ gauge theory.
In section 5 we discuss generalizations of this construction; this
includes generalizations to $c=1$ non-critical string at multiples of
self-dual radius and the decompactified limit.  We show how the
decompactified $c=1$ bosonic string theory can be viewed as computing
$F$-terms for certain deformation of $(2,0)$ superconformal theory in
6 dimensions.  In section 6 we discuss application of deconstruction
to arbitrary $k$-dimensional bosonic gauge theory and its
interpretation in 4-dimensional terms, including deconstruction of
pure $\cN=1$ Yang-Mills in 5 dimensions.  In section 7 we discuss how
these ideas may lead to the full reconstruction of string theory in
certain backgrounds from matrix models.

\newsec{$c=1$ Non-Critical String Theory}

In the continuum formulation the world-sheet theory of the $c=1$
string consists of a single bosonic coordinate $X$ that can be
compactified on a circle of radius $R$
$$
X \sim X + 2\pi R.
$$
Because this string is non-critical ($c\not=26$) the conformal mode of
the world-sheet metric does not decouple and becomes a second
dynamical space-time coordinate $\v$, the Liouville field, making the
target space effectively two-dimensional. The local world-sheet
Lagrangian reads
$$
\int d^2z\left( \hf (\d X)^2 + (\d\v)^2 + \m \,e^{\sqrt{2}\v}
+ \sqrt{2} \v R^{(2)} \right)
$$
together with a pair of $(b,c)$ diffeomorphism ghosts. Here the
background charge for $\v$ gives total central charge $c=26$ for the
matter fields. The coupling $\mu$ plays the role of a world-sheet
cosmological constant. Its presence makes the CFT strongly interacting
and difficult to analyze, although for specific amplitudes the
Liouville potential can be treated perturbatively in $\mu$.

\subsec{Physical States and The Ground Ring}

At arbitrary compactification radius $R$ the physical states of this
string theory are mainly of two types. (Note that all operators are
dressed by an appropriate Liouville vertex operator $e^{s\v}$ to give
total scaling dimensions $(1,1)$.)  First, there are the tachyon
momentum operators
$$
T_k = e^{ikX/R}
$$
that create the modes of the two-dimensional  massless ``tachyon''
field $T(X,\v)$. Furthermore there are the winding modes
$$
\tT_m = e^{im\tX R}
$$
with $\tX$ is the dual scalar field, obtained by a T-duality from $X$,
that is compactified on a circle of radius $1/R$. There are no mixed
momentum/winding operators, since physical vertex operators should
have equal left and right conformal dimensions $h=\overline h$.  Apart
from these tachyon modes there are also so-called discrete states \kp,
that are most relevant at the self-dual radius.

At the self-dual radius $R=1$ the $U(1)\times U(1)$ symmetry of the
$c=1$ conformal field theory is extended to a $SU(2)\times SU(2)$
affine symmetry. Under this extended symmetry the vertex operators
$T_k$ become part of a spin $(k/2,k/2$) multiplet of primary fields
$$
\cO^{(k)}_{n_l,n_R},\qquad -k \leq n_L,n_R \leq k.
$$
The highest/lowest weight operators are related to the tachyon vertex
operators as
$$
\cO^{(k)}_{\pm k,\pm k} = T_{\pm k},\qquad
\cO^{(k)}_{\pm k,\mp k} = \tT_{\pm k}.
$$

These physical states satisfy interesting algebraic relations. In
any string theory physical vertex operators can be represented in
two pictures: either as BRST-closed 0-forms on the world-sheet, or
as (1,1)-forms that can be consistently integrated over the the
world-sheet. In the former representation there is a natural ring
structure, given by the operator product (modulo BRST exact terms)
--- the so-called ground ring. In the correspondence with twisted
$\cN=2$ superconformal field theories this ground ring can be
identified with the chiral ring.

As observed by Witten \witg\ in this case this ring has a simple
geometric structure. Introduce two doublets $(a_1,a_2)$ and
$(b_1,b_2)$ of the $SU(2) \times SU(2)$ symmetry group. Then the
representations $\cO^{(k)}$ of spin $(k/2,k/2)$ is given by
expressions $P(a)Q(b)$ where $P(a)$ and $Q(b)$ are polynomials of
degree $k$. Stated otherwise, we can write the basis of string
observables as
\eqn\Ostring{
\cO^{(k)}_{n_L,n_R} = a_1^{n_L}a_2^{k-n_L}
b_1^{n_R}b_2^{k-n_R}.
}
The ground ring is captured by introducing the four generators
$$
x_{ij} = a_i b_j.
$$
In the representation as $(1,1)$ forms they can be identified as
gravitationally dressed versions of the minimal momentum/winding
operators
$$
\eqalign{
x_{11} = T_{+1},\qquad & x_{12} = \tT_{+1}, \cr
x_{21} = \tT_{-1},\qquad & x_{22} = T_{-1}. \cr
}
$$
The generators $x_{ij}$ satisfy a single relation that at $\m=0$ reads
$$
\det x_{ij} = x_{11} x_{22} - x_{12} x_{21} = 0.
$$
When considered with complex variables this affine quadric defines the
conifold --- a singular Calabi-Yau threefold. As has been argued in
\witg, in the $c=1$ string with $\m\not=0$ this relation is
generalized to
\eqn\deform{
x_{11} x_{22} - x_{12} x_{21} = \m,
}
which is the real version of the deformed conifold.
Viewing $x_{ij}$ as complex variables would lead to
a geometry diffeomorphic (as a
real 6-manifold) to $T^*S^3$.
This is not an accident.  In fact as shown in \ghv\
the topological B-model on the deformed conifold is equivalent
to $c=1$ non-critical strings at the self-dual radius.
Turning on the general deformations
$\cO^{k}$ will deform this geometry to a general affine
hypersurface
\eqn\defor{
x_{11} x_{22} - x_{12} x_{21} + f(x_{11},x_{22},x_{12},x_{21})=\mu.
}
This geometry can be seen as a general local CY geometry in the
neighbourhood of a (deformed) conifold singularity. In that sense the
$c=1$ string captures the data of the general topological B-model on a
local CY three-fold. Note that by the Morse lemma we can always put a
local geometry of the form \defor\ {\it locally} in the canonical form
\deform. But of course such a transformation changes dramaticaly the
behaviour at infinity, and therefore changes the physics.

\newsec{Affine Quiver Qauge Theories and Supermatrix Models}

The $\cN=1$ gauge theory that we will be discussing in this section
has been investigated in great detail, because it has a holographic
dual given by IIB string theory on $AdS_5 \times T^{1,1}$ with fluxes
\kw. Let us briefly recall the construction of the gauge theory and
how it is obtained from D-branes in a conifold geometry.

\subsec{The Supersymmetric Gauge Theory}

One starts with a compactification of the IIB string theory on the
singular conifold
$$
x_{11} x_{22} - x_{12} x_{21} = 0.
$$
One then puts $N$ D3 branes at the singularity. Their world-volumes
completely fill the space-time $\R^4$. This brane configuration gives
in the near-horizon or decoupling limit a superconformal quiver gauge
theory with gauge group $U(N) \times U(N)$ \kw.

Before recalling the field content of this gauge theory, it is natural
to consider a more general class of models. One can break the
conformal symmetry by placing $K$ additional D5 branes that wrap the
$\P^1$ obtained by a small resolution of the conifold.  In that case
we obtain a $\widehat{A}_1$ quiver gauge theory \refs{\kg,\kt} with
gauge group
$$
U(N_+) \times U(N_-),
$$
with the ranks of the two gauge groups given as
$$
N_+=N+K,\qquad N_-=N.
$$
Furthermore this gauge theory contains the following chiral matter
fields. There are two fields $\F_+$ and $\F_-$ in the adjoint
representation of $U(N_+)$ and $U(N_-)$ respectively. These are
supplemented by two sets of bi-fundamental fields: $A_1,A_2$
transforming in the representation $(\overline{N}_+,N_-)$, and
$B_1,B_2$ transforming in the representation
$(N_+,\overline{N}_-)$. There is an obvious $SU(2)
\times SU(2)$ global symmetry acting on these bi-fundamentals, with
the $A_i$ transforming as one doublet and the $B_i$ as another.

The tree-level superpotential for these fields is given by \kw
\eqn\superpot {
W = {1\over 2}\left(\Tr\, \F_+^2 -
\Tr\,\F_-^2\right) - \Tr\,A_i\F_+ B_i + \Tr\,B_i\F_- A_i.}
Integrating out the adjoint scalars produces the quartic
superpotential
\eqn\wkw{
W = {1\over 2} \Tr\left(A_1B_1A_2B_2 - A_1B_2A_2B_1\right).
}
At the critical point of $W$ we find the relations
$$
\F_+ = B_iA_i,\qquad \F_- = A_iB_j,
$$
and
$$
A_i \F_+ = \F_- A_i , \qquad \F_+ B_i = B_i \F_- .
$$
These relations allow us to express the adjoints $\F_\pm$ in terms of
the bifundamentals $A_i,B_i$. Furthermore, the latter satisfy
relations that can be written as
\eqn\exchange{
A_1 B_j A_2 = A_2 B_j A_1,\qquad B_1 A_j B_2 = B_2 A_j B_1.
}

A basis of chiral operators of this gauge theory is given by
expressions of the form
\eqn\Ogauge{
\cO^{(k)}_{i_1\ldots i_k,j_1,\ldots j_k}
= \Tr\(A_{i_1} B_{j_1}\cdots A_{i_k} B_{j_k}).
}
Because of the relations \exchange\ the tensor $\cO^{(k)}$ is
completely symmetric in the $i$ and the $j$ indices. In terms of the
$SU(2) \times SU(2)$ global symmetry these operators therefore
transform as a spin $(k/2,k/2)$ representation. In the superconformal
point these fields have scaling dimension $3k/2$ and R-charge
$k$. Besides these fields there are the corresponding combinations
including the $U(N_+) \times U(N_-)$ glueball superfield $\cW_\a$
$$
\Tr\left[\cW_\a (AB)^k\right],\quad
\Tr\left[\cW_\a \cW^\a (AB)^k\right].
$$
All these chiral fields have been identified under the holographic
duality to IIB string theory on $AdS_5 \times T^{1,1}$
\ceresole.

The similarity of this set of gauge theory observables \Ogauge\ to the
observables \Ostring\ of $c=1$ at self-dual radius \witg\ was noticed
in this context by \rand. As we will explain
below this is not accidental.

\subsec{The Cascade}

As explained in \klebs\ the dynamics of this gauge theory is extremely
rich.  Under the RG flow alternatively the gauge coupling of one of
the two gauge groups will go to strong coupling, and the theory will
undergo a Seiberg-duality \seiberg.  There is also another version of
description of this duality, discussed in \cfikv\ which one keeps the
adjoint fields and the cascade is described as an affine Weyl
reflection.  This description of the duality can be immediately
implemented in the matrix model. We will discuss this later in the
context of matrix model.  Here we will review the Seiberg duality as
applied to this case in \klebs.

Suppose that $N_+ > N_-$ and that the gauge group $U(N_+)$ becomes
strongly coupled as we go towards the IR. So we will perform a Seiberg
duality with $N_+$ colors and $2N_-$ flavors.  The duality will then
replace the strongly coupled gauge group as
$$
U(N_+) \to U(2N_- - N_+)
$$
and replace the bifundamentals $A_i,B_j$ by new bifundamentals
$a_i,b_j$. Apriori there is no relation between these fields.
Furthermore there are new meson fields
$$
M_{ij} = A_i B_j
$$
where we have contracted the $U(N_+)$ indices. So the mesons $M_{ij}$
are neutral with respect to $U(N_+)$ but transform as adjoints under
the `flavor gauge group' $U(N_-)$. There is further a tree-level
superpotential
$$
\Tr(M^{ij} a_i b_j).
$$
(Here $SU(2)$ indices are raised and lowered using $\e_{ij}$).
Plugging in these fields in
the original Klebanov-Witten superpotential \wkw\ gives
$$
W= \Tr\left(M_{11} M_{22} - M_{12} M_{21} + M^{ij} a_i b_j\right)
$$
It is imple to integrate out the meson fields, that only appear quadratically.
After a shift
$$
M_{ij} \to M_{ij} + a_i b_j
$$
we obtain the dual superpotential
$$
W = \Tr\left(a_1b_1 a_2 b_2 - a_1 b_2 a_2 b_1 \right).
$$
This is the same quartic superpotential now written in terms of the
new bifundamentals for the $U(2N_- -N_+) \times U(N_-)$ gauge theory.

We now want to see what happens to this argument when we have
deformations of $W$ by arbitrary monomials in the $A_i,B_j$
$$
W= \Tr(A_1B_1A_2B_2-A_1B_2A_2B_1) + \sum t_{i_1\cdots i_k,j_1\cdots j_k}
\Tr(A_{i_1}B_{j_1}\cdots A_{i_k}B_{j_k}).
$$
Let us keep the variation of $W$ infinitesimal. That is, we will be
interested in computing correlation function of chiral operators in
the undeformed theory with the couplings $t=0$. In
this case the cascade goes through and we get the desired result by
replacing the monomials in terms of the meson fields and then
eliminating meson fields.

So, for example, in this way the operator $\Tr(A_1B_1)^k$ gets replaced
by $\Tr M_{11}^k$. One then has to eliminate the meson fields using
the above superpotential. To leading order the equations for the
elimination of meson fields do not get modified, and we can simply replace
$$
M_{ij} \to a_ib_j.
$$
This means that we have the same set of chiral fields, but now written
in terms of the new bifundamentals $a_i,b_j$. The same chiral operator
now takes the form $\Tr(a_1b_1)^k$. But beyond the leading order in
the couplings $t_k$, it does change the basis of chiral fields. This
implies that in the process of the Seiberg duality we will have
operator mixing among the chiral fields.

A particular interesting case appears if we start with a gauge theory
with rank
$$
N_+=2N, \qquad N_- = N.
$$
If we apply the duality to this case, we are left with a pure $U(N)$
gauge theory. The other gauge factor has disappeared.  So there are no
longer new bifundamental fields $a_i,b_j$. The only dynamical fields
are the mesons $M_{ij}$ that we recall transform as adjoints under the
remaining gauge group $U(N)$. This critical case with $N_c=N_f$ is
more subtle, since there are now baryon degrees of freedom, and a
well-known quantum correction to the moduli space.  Ignoring these
effects, the superpotential for this model, including arbitrary
deformations will be of the form
\eqn\mmm{
W = \Tr\left(M_{11} M_{22} - M_{12} M_{21} + f(M_{ij})\right)
}
with $f$ an arbitrary function of the four meson fields $M_{ij}$.  We
will see in a moment how, using the relation of F-term computations in
gauge theory to matrix models, this suggests a four-matrix model for
this $U(N)$ gauge theory with four adjoint chiral fields.

Clearly, in the connection with the $c=1$ string we want to identify
the mesons fields $M_{ij}$ with the space-time coordinates $x_{ij}$ of
the conifold, and relate the superpotential deformation \mmm\ to the
deformed conifold \defor.  Note also that the cascade at the level of
gauge theory has been interpreted as flops in \cfikv\ and thus at
the topological level ({\it i.e.} the theory on topological branes)
the cascade continues to hold for the matrix model.

\subsec{The Quiver Matrix Model}

According to the results of \refs{\dvi,\dvii,\dviii} the effective
superpotential of the quiver gauge theory can be computed in terms of
an associated quiver matrix model. More precisely, the effective
superpotential, considered as a function of the glueball superfields
$S_+$ and $S_-$ associated to the $U(N_+)$ and $U(N_-)$ gauge groups,
takes the form
$$
W_{\rm eff}(S) = N_+ {\d \cF_0 \over \d S_+} +
N_- {\d \cF_0 \over \d S_-}+ 2\pi i (\t_+ S_+ + \t_- S_-),
$$
where the function $\cF_0(S_+,S_-)$ is the planar free energy of the
associated matrix model, and $\tau_\pm$ are the bare gauge couplings
of the two gauge factors.

The corresponding random matrix model consists of the same field
content as the gauge theory, but the ranks $M_+,M_-$ of the matrices
$\F_+,\F_-$ are unrelated to the ranks $N_+,N_-$ of the gauge
groups. In the 't Hooft limit $g_s\to 0$, $M_\pm \to \infty$, we
instead identify
$$
S_\pm= g_s M_\pm.
$$
The matrix integral can be written as
$$
Z = {1\over V}\int d\F_+ d\F_- dA_i dB_i \exp
\left[-{1\over g_s} W_{\rm tree}(\F_+,\F_-,A_i,B_j)\right]
$$
with normalization factor
$$
V=\vol\left(U(M_+) \times U(M_-)\right)
$$

We now want to claim that the matrix model only depends on the
combination
$$
S= S_+ - S_-.
$$
In particular we have a simple relation between the matrix model free
energy $\cF_0(S)$ and the gauge theory effective superpotential
\eqn\weff{
W_{\rm eff}(S) = (N_+ - N_-) {\d \cF_0 \over \d S} + 2\pi i (\t_+
-\t_-)S.
}
We will show that the free energy $\cF_0(S)$ is just that of the
gaussian one-matrix model.

To analyze the model we can work with a more general superpotential
\eqn\gen{
\Tr\Bigl(W(\F_+) - W(\F_-) - A_i\F_+B_i + B_i\F_-A_i\Bigr).
}
As a first step, one can integrate out the bifundamentals $A_i,B_j$
and then go to an eigenvalue basis for the remaining adjoint fields $$
\F_\pm \sim {\rm diag}\left(\l_1^\pm,\ldots,\l^\pm_{M_\pm}\right).
$$
This reduces the matrix model to the following integral over the
eigenvalues
\eqn\eigenvalues{
\eqalign{
Z & = \int \prod_{I,K} d\l_I^+ d\l_K^-
\ {\displaystyle \prod_{I<J} \left(\l_I^+ -\l_J^+\right)^2
\prod_{K<L} \left(\l_K^- -\l_L^-\right)^2 \over \displaystyle
\prod_{I,K} \left(\l_I^+ - \l_K^-\right)^2} \cr &
\qquad \qquad \qquad \times \
\exp \left[-{1\over 2g_s}\left(\sum_I W(\l_I^+)
- \sum_K W(\l_K^-)\right)\right]}}

The integral \eigenvalues\ represents a gas of both positively and
negatively charged eigenvalues with a Coulomb interaction in a
background potential $W(x)$. To find the large $M$ saddle-point, it
helps to consider it as a system of a total of $M_+ + M_-$ eigenvalues
$\l_I$ that each can have a charge $q_I=\pm 1$. There will then be
$M_+$ eigenvalues of positive charge and $M_-$ eigenvalues of negative
charge. If one wishes, the negatively charged eigenvalues can be
considered as ``holes'' or ``anti-eigenvalues.''

With this notation we can write the integral as
$$
Z = \int \prod_{I} d\l_I \ \prod_{I<J} \left(\l_I
-\l_J\right)^{2q_I q_J} \exp \left[-{1\over g_s}\sum_I q_I
W(\l_I) \right]
$$
The equation of motion now reads
$$
W'(\l_I) = 2g_s \sum_{J\not=I} {q_J \over \l_I - \l_J}.
$$
The saddle-point evaluation of this integral in the 't Hooft limit
proceeds now very much like the bosonic one-matrix model with action
$\Tr\, W(\F)$. One introduces the resolvent
$$
\w(x) = {1\over M} \sum_I {q_I \over x - \l_I},
$$
with $M$ the net charge in the system
$$
M = \sum_I q_I = M_+ - M_-.
$$
We similarly denote $S =g_s M = S_+ - S_-.  $ One then
straightforwardly derives the planar loop equation
\eqn\spectral{
y^2 - W'(x)^2 + f(x)=0
}
written for the variable
\eqn\force{
\eqalign{
y & = W'(x) - 2S \w(x) \cr & = W'(x) - 2g_s \sum_I {q_I \over x -
 \l_I}.  }}
Here the quantum deformation $f(x)$ is given as the weighted average
$$
f(x) = 4g_s \sum_I q_I {W'(x) - W'(\l_I) \over x - \l_I}.
$$
For a general potential $W(x)$ of degree $n+1$ this is a polynomial of
degree $n$, that encodes the dependence on the moduli $S_i=g_s
M_i$. Here the relative filling fractions $M_i$ now indicate the total
charge $\sum q_I$ of the eigenvalues that occupy the $i$-th critical
point of $W$.

We conclude that the spectral curve \spectral\ of the $\widehat{A}_1$
quiver matrix model is identical to that of the bosonic matrix model
\dvi, with the remark that the filling fractions $M_i$ and therefore
also the moduli $S_i$ now also can be naturally negative. In fact, if
we take a real potential $W(x)$ and hermitian matrices $\F_\pm$, then
the eigenvalues of positive charge will typically sit at the (stable)
minima, while those of negative charge will sit at the (unstable)
maxima.

Note that the $B$-cycle period
$$
{\d \cF_0 \over \d S_i} = \int_a^\infty y dx
$$
have an interpretation as either removing a positively charged
eigenvalue from the system, or adding a negatively charged one.  The
system is therefore insensitive to adding an
eigenvalue/anti-eigenvalue pair.  So the planar free energy $\cF_0$ of
quiver matrix model, as a function of the variables $S_i$, is exactly
the same as that of the one-matrix model. In fact, since the full loop
equations for $\w(x)$ are identical to those of the one-matrix model,
this identity holds also for the higher genus contributions $\cF_g$.

In the special case of the gaussian potential $W(x) = \hf x^2$ the spectral
curve of the affine quiver is
$$
y^2 + x^2 = S
$$
and the planar free energy is given by
$$
\cF_0(S) = \hf S^2 \log(S/\L^3).
$$
Plugging this into \weff\ we get the effective superpotential for the
quiver system
$$
W_{\rm eff}(S)  = (N_+ - N_-) S \log(S/\L^3) + 2\pi i (\t_+
-\t_-)S.
$$

\subsec{Supermatrix Models}

As an aside we like to point out that integrals like the quiver matrix
model \gen\ have been considered before in the context of supermatrix
models (here defined as integrals over super Lie algebras, see {\it
e.g.}\ \supermatrix). In that case we work with a $M_+|M_-$
supermatrix $\F$ with a decomposition
$$
\F = \left(\matrix{\F_+ & \psi \cr \chi & \F_- \cr}\right),
$$
with $\F_+,\F_-$ bosonic and the off-block diagonal components
$\psi,\chi$ fermionic. The action is now written in terms of a
supertrace as
$$
{\rm Str}\, W(\F), \quad \hbox{with}\quad 
{\rm Str}\left(\matrix{a & b \cr c & d \cr}\right) = a -d.
$$
This action is invariant under the supergroup $U(M_+|M_-)$. It has been
observed that this supermatrix model is equivalent to the $U(M)$
bosonic matrix model with $M=M_+ - M-$. This result is essentially
equivalent to invariance under the duality cascade.  Following
\vafadd\ we expect this system to appear naturally from the
topological field theory description a sytem of $M_+$ D5 branes and
$M_-$ anti-D5 branes together with a set of D3 branes represented as
non-trivial flux.

As in \dgkv\ we can go to a basis in which the odd components are zero
$$
\F = \left(\matrix{\F_+ & 0 \cr 0 & \F_- \cr}\right),
$$
and we break to the bosonic subgroup
$$
U(M_+|M_-) \to U(M_+) \times U(M_-).
$$
This introduces ghosts $b,c$ with a decomposition
$$
b = \left(\matrix{0 & B \cr A & 0 \cr}\right),\qquad c =
\left(\matrix{0 & B^* \cr A^* & 0 \cr}\right).
$$
Because the ghosts have odd statistics the off-diagonal fields are
even; we will decompose them as
$$
A,A^*=A_1 \pm i A_2,\qquad B,B^*=B_1 \pm i B_2.
$$
The matrix model action in the gauge fixed version is given by a
supertrace
\eqn\wgauge{
W= {\rm Str}\left(W(\F) + b[\F,c]\right).
}
which written in components reduces exactly to \gen.

\subsec{Stability of The Glueball Superpotential and K-Theory}

The above ideas are relevant for the resolution of the following
puzzle: If we fix a gauge group of finite rank $N$, the powers of the
glueball superfield $S$, defined as a fermion bilinear, terminate at
finite order, if we consider $S$ as a classical field. How can one
then justify the perturbative computations of the glueball
superpotential, as was done successfully in \dglvz\ not incorporating
this fact?  One resolution of this is to use the ``replica trick'' and
embed the $U(N)$ theory in the $U(NK)$ theory, and do the computation
in that context in the limit where $K$ is large.  But under this
transformation the superpotential changes by a factor of $K$, and so
in particular one is dealing with a different effective
theory. Moreover, one will have to argue why going back to the $U(N)$
theory is as simple as dividing the superpotential by a factor of $K$.

We feel a better resolution of this puzzle is what we have observed
here: The F-term of the supersymmetric gauge theory is {\it
equivalent} to that of a bigger gauge theory by an inverse cascade
effect.  This, as we have explained above, is related to adding $M$
extra brane/anti-brane pairs to an ${\cal N}=1$ theory with an
equivalent F-term content. In the context of this bigger gauge theory,
the computation of the glueball superpotential makes sense for higher
powers of $S_+$ and $S_-$, as long as we take $M$ large enough.  In
fact we can take $M\rightarrow \infty$ without changing the F-terms of
the theory, and consider arbitrary powers of $S_\pm$'s since the
ranks become infinite. This in particular justifies the perturbative
computation done in \dglvz, without needing to change the F-term
content of the theory.

This is very similar to how K-theory captures D-brane charges, where
one considers adding an arbitrary number of brane/anti-brane pairs
\witk .  We thus see a notion of ``stability'' of the perturbative
glueball superpotential computation, where one stabalizes the gauge
theory by adding sufficient numbers of branes and anti-branes so that
one can effectively ignore the condition that the glueball field $S$
is nilpotent classically.

\subsec{Seiberg-like Dualities From Matrix Models}

Quiver theories with gauge group $\prod U(N_i)$, consisting of adjoint
chiral field $\Phi_i$ on each node with some superpotential $W_i
(\Phi_i)$ and certain bifundamental fields $Q_{ij}$, admit a duality
discovered in \cfikv, considered in the context of affine A-D-E quiver
theories. We replace the rank of the gauge group at node $i$ by the
sum of the adjacent ranks minus $N_i$ (this is the analog of $N_f-N_c$
for Seiberg duality), and we replace
$$W_i(\Phi_i)\rightarrow -W_i(\Phi_i)$$
 and
$$W_j(\Phi_j)\rightarrow W_j(\Phi_j)+e_j\cdot e_i W_i(\Phi_j),$$
where $e_i$ denote the basis of positive roots associated to the nodes
of the affine quiver theory. This duality was interpreted in \cfikv\
as a Weyl reflection on the node $i$.  Exactly the same interpretation
can be done in the context of matrix model, which gives a derivation
of this duality; this point was already noted in \dviii\ and we will
elaborate on it here.

The matrix model which describes the F-terms of this theory is the
quiver matrix model already studied in \kostov.
One can integrate out the bifundamental fields (as we did above for
the case of the $\widehat{A}_1$) to obtain the integral in term of the
eigenvalues of the $\Phi_i$ on each node:
$$Z=\int \prod_{i,I} d\l^i_I\ \prod_{(i,I) \not=(j,J)}
\left(\lambda^i_I -\lambda^j_J\right)^{e_i\cdot e_j}
\exp\sum_{i,I} W_i(\lambda^i_I)
$$
This system is clearly invariant under the Weyl reflection.  To
make this more clear, consider the planar limit, where we associate
a density eigenvalue $\rho^i(\lambda)$ to each node $i$.  Then the
effective action can be written as
$$
S=\int  d\lambda\  \rho^i (\lambda) W_i (\lambda)+
\int d\lambda d\lambda'\ (e_i\cdot e_j) \rho^i(\lambda) \rho^j(\lambda')
\log(\lambda -\lambda ')$$
Viewing $\rho =\rho^i e_i$ and $W=W_i e^i$ as vectors and covectors in
the affine root lattice this system enjoys the natural Weyl reflection
action on each node, which leads to the duality mentioned above.  As
was noted in \cfikv\ this leads to a matrix model derivation of the
gauge theory duality\foot{In fact one can refine the statement of the
duality, as was done in \cfikv, to explain how the critical points of
the matrix model (which correspond to arbitrary {\it roots} of the
affine Dynkin diagram) get exchanged under the Weyl reflection.}.

\newsec{$c=1$ Non-Critical Bosonic String and $\cN=1$ Gauge Theory}

So far we have seen that a particular gauge theory, namely the affine
quiver theory based on ${\widehat A}_1$ with quadratic superpotential,
has the same set of chiral fields as that of the $c=1$ string at
self-dual radius. This is not accidental, as we will explain in this
section. Furthermore we give the detailed link between computations
of $c=1$ correlation functions and F-terms of $\cN=1$ supersymmetric
gauge theory. This will generalize the correspondence of the free
energies to arbitrary correlators.

It was argued in \ghv\ (using the result of \mv\ describing $c=1$ at
self-dual radius as a $SL(2)/U(1)$ Kazama-Suzuki model at level $3$)
that B-model topological string theory on the deformed conifold
geometry
$$x_{11}x_{22}-x_{12}x_{21}=\mu$$
is equivalent to non-critical bosonic strings with $c=1$ at self-dual
radius, where one identifies the geometry as the (complexified) ground
ring of the bosonic string theory. This identification was checked to
a few loop orders and extended to all loops using the results
\nareatal. This identification was further studied in \oogvaf.

However, this topological model is equivalent (as a mirror of the duality
between Chern-Simons and topological strings \gv ) to the resolved
conifold geometry with branes wrapping $\P^1$. In the connection
between the topological string and the type IIB string, these branes
get promoted to $D5$ branes, and in addition $D3$ branes, which are
points in the internal geometry. On the other hand the gauge theory of
this system involves the ${\widehat A}_1$ quiver theory \kw\ already
discussed. Since the B-model topological string on the side including
branes computes $F$-terms of the corresponding $\cN=1$ gauge theory,
and that is equivalent to the matrix integral, we come to the
conclusion that the $c=1$ non-critical bosonic string at self-dual
radius is equivalent to the ${\widehat A}_1$ quiver matrix model. We
explain this identification now in more detail.

As discussed before for each observable of the $c=1$ theory at self
dual radius, there is a chiral field in the matrix model and the gauge
theory, namely
$$
\cO^{(k)}_{i_1...i_k,j_1...j_k}\ \leftrightarrow \ \Tr(A_{i_1}B_{j_1}...
A_{i_k}B_{j_k})
$$
Consider now the partition function of $c=1$ theory deformed by
arbitrary physical operators
$$
Z(\mu,t)=
\Bigl\langle \exp \sum t_{i_1\cdots i_k,j_1\cdots j_k}
\cO^{(k)}_{i_1\cdots i_k,j_1\cdots j_k} \Bigr\rangle
$$
with genus expansion
$$
Z(\m,t) = \exp \sum_{g \geq 0} g_s^{2g-2} \cF_g(\m,t).
$$
Then we reach the conclusion that this is equivalent to the deformed
${\widehat A}_1$ matrix model theory
\eqn\corr{
\eqalign{
Z(\mu,t) =& {1\over V}
\int dA_idB_j\ \exp {1\over g_s}\Bigl[ \Tr(A_1B_1A_2B_2-A_1B_2A_2B_1) \cr
& \qquad\qquad + \sum t_{i_1\cdots i_k,j_1\cdots j_k}
\Tr(A_{i_1}B_{j_1}\cdots A_{i_k}B_{j_k}) \Bigr] \cr }}
where $\mu =g_s(M_+-M_-)$ ($S$ in the gauge theory) and $M_+$ and
$M_-$ are the ranks of the two matrices of ${\widehat A}_1$ theory.
As discussed before the identification of the parameters
$t_{i_1...i_k, j_i...j_k}$ is unambiguous to first order but beyond
that there could be an operator mixing.

Moreover if ${\cal F}_0(\mu,t)$ denote the contribution of planar
diagrams (or equivalently the genus $0$ amplitudes of $c=1$ at
self-dual radius) then in the associated $\cN=1$ supersymmetric gauge
theory the deformed superpotential is given by
$$
W_{\rm eff}(S,t) = (N_+-N_-){\d\cF_0(S,t)\over \d S} + (\t_+ - \t_-) S.
$$

Using the cascade the associated gauge theory can also be viewed
as a single $U(N)$ gauge theory with 4 adjoint fields $M_{ij}$.
Ignoring the subtleties of baryons and the quantum moduli space, the
naive superpotential for the associated 4-matrix model is
$$
W=\Tr\left({\rm det}_{ij} M  +f(t,M)\right).
$$
In particular if we just deform the momentum modes of the $c=1$
self-dual string, {\it i.e.} if we want to compute the scattering
amplitudes of the tachyon modes
$$
Z(\mu,t)= \Bigl\langle \exp \sum_k t_k T_k\Bigr\rangle,
$$
two adjoint fields can be integrated out. So the correlations get
mapped to an $\cN=1$ supersymmetric gauge theory with the
remaining two adjoints 
$$
X_+=M_{11},\qquad X_-=M_{22},
$$
and with superpotential
\eqn\twomm{
W=\Tr\(X_+X_- +\sum_{n>0} {t_n} X_+^n+{t_{-n}} X_-^n\).
}
This is exactly a well-known two-matrix model representation of the
$c=1$ string \normal.  Using the Harish Chandra-Itzykon-Zuber
integral, one can show that this matrix model is equivalent to another
model where one assumes that the matrices $X_+,X_-$ commute. Since we
can also choose the reality condition $X_+^\dagger=X_-$, this is
sometimes also called the normal matrix model\foot{A normal matrix $X$
satisfies $[X,X^\dagger]=0$.}.

A generalization of this model that includes couplings $\tt_k$ to the
winding modes or vortices $\tT_k$ has been proposed in \kazakov. In
this three-matrix model, one also introduces a unitary matrix $U$
that captures the vortices. The full action is
\eqn\kaz{
 \Tr\(X_+ X_- - UX_+ U\inv X_-
+ \sum_{n>0} (t_{+n} X_+^n + t_{-n} \Tr\, X_-^n)
+ \sum_{k \in \Z} \tt_k \Tr\,U^k\).}
It would be very interesting to relate this models directly to our
quiver model. (See in this respect also our comments in section 6.)

\newsec{Generalizations}

The above example can be generalized in two basic directions. On the
one hand we can generalize this to the case where the radius of the
$c=1$ circle is $k$ times the self-dual radius. On the other hand, as
we discussed before, we can relax the condition of the adjoint
superpotential to be quadratic.  We will consider these two cases in
turn.

\subsec{Multiples of The Self-Dual Radius}

The fastest way to obtain the matrix model and the gauge theory in
this case is to consider a $\Z_k$ orbifold of the model with self-dual
radius --- on the side of both the bosonic string, the matrix model,
the gauge theory and the geometry. Let us consider these in turn.

Recall that in the case of the non-critical $c=1$ bosonic string at
self-dual radius the Calabi-Yau 3-fold geometry can be identified with
the ground ring.  In particular, the ring at zero cosmological
constant
$$x_{11}x_{22}-x_{12}x_{21}=0$$
is identified with the singular conifold. Recall also that the
monomials are identified with the momentum and winding modes of $c=1$
as discussed in section 2.1. Thus under the $\Z_k$ orbifold they
transform according to
$$
\eqalign{
x_{11}& \rightarrow \omega x_{11},\cr
x_{22}& \rightarrow \omega^{-1}x_{22},\cr
x_{12} & \rightarrow x_{12},\cr
x_{21} &\rightarrow x_{21},\cr}
$$
where $\omega$ is a primitive $k$-th root of unity
$$\omega^k=1.$$
The invariant ring in this case is generated by
$$u=(x_{11})^k,\qquad v=(x_{22})^k,$$
together with $x_{12},x_{21}$. This leads to the ring relation
\eqn\rk{uv=(x_{12} x_{21})^k}
This is indeed the chiral ring for $k$ times the self-dual radius
\muet.

One can also carry this orbifolding at the level of the quiver theory.
This has been done in \ura. One ends up with an ${\hat
A}_{2k-1}$ quiver theory with $N=2$ matter content and bifundamental
fields between the nodes. Moreover there is a superpotential
$+\Tr\,\Phi_i^2$ for the adjoint field for the even nodes and
$-\Tr\,\Phi_i^2$ for the odd nodes. To obtain this orbifold one simply
uses the method of obtaining quiver theories on D-brane orbifolds
\dm\ where the $\Z_k$ action on the fields is given,
in addition to the cyclic action on the $k M_+$ and $k M_-$ branes, by
$$
\eqalign{
A_1 & \rightarrow \omega A_1,\cr
B_1 & \rightarrow B_1, \cr
A_2 & \rightarrow A_2, \cr
B_2 & \rightarrow \omega^{-1} B_2, \cr
\Phi_+ & \rightarrow \Phi_+, \cr
\Phi_- & \rightarrow \Phi_-. \cr
}
$$

We can also obtain an alternative derivation of the quiver theory
starting from the ring relations \rk\ and using the results of \ckv\
and \gn. Let us define $x_{12}=y+x$ and $x_{21}=y-x$.  Then the
relation \rk\ becomes
$$uv-(y^2-x^2)^k=0,$$
which can be equivalently written as
$$uv-(y-x)(y+x)(y-x)\cdots(y+x)=0.$$
Here there are $k$ monomials of $(y+x)$ and $k$ monomials of $y-x$.
According to \ckv\ if one has a geometry of the form
$$uv-\(y-e_1(x)\)\(y-e_2(x)\)...\(y-e_{2k}(x)\)=0,$$
with branes wrapped over blown up 2-cycles, one obtains the
${\widehat A}_{2k-1}$ quiver theory with $\cN=2$ content
where the gradient of the superpotential as a function of the
$i$-th adjoint field is
$$W_i'(x)=e_i(x)-e_{i+1}(x),$$
with $x\rightarrow \Phi_i$. Moreover the sum of the $W_i$'s are zero.

Applying this to the case at hand we see that we have an adjoint field
with quadratic superpotential at each node, with alternating signs for
nearest neighbors. Thus we have a reformulation of the $c=1$ bosonic
string at $k$ times the self-dual radius in terms of a matrix quiver
theory.  Note that there are more gauge groups and we can vary the
rank of all $2k$ gauge groups.  In particular we can choose all the
ranks $M_i$ (with $i=1,\ldots,2k$) independently.  This corresponds in
the $c=1$ theory to turning on twisted states, which are the lightest
momentum modes in the theory with radius being $k$ times the self-dual
radius.

\subsec{More General Superpotentials}

In principle we can have an independent superpotential for each of the
nodes of the affine quiver theory as long as they sum up to zero
\ckv. For example for the ${\widehat A}_1$ quiver theory, as we already
explained, we could put superpotentials $W(\Phi_+) -W(\Phi_-)$ for a
general polynomial $W$. In this case the matrix model computations
reduce to that of the single matrix model with potential $W(\Phi)$.
Note that this can be viewed as a deformation of the quadratic
potential by some higher powers of $\Phi$ which can be written in
terms of monomials of $A_i$ and $B_j$ using
$$
\Tr\,\F_+^n = \Tr(A_iB_i)^n.
$$
Thus the correlations of the gaussian matrix model observables $\Tr\
\F^n$ can be viewed as computing some specific subset of correlation
functions of $c=1$ at self-dual radius.

This relation can also be seen from the geometry side. Including
higher powers of $\F$ deforms the conifold to
$$
uv+y^2 + W'(x)^2 + f(x)=0
$$
This is a special case of the universal deformation \defor.

Similarly we can consider the $\Z_k$ orbifold of this and obtain
alternating $W$'s at each node.

\subsec{Deconstruction and ${\widehat A}_\infty$ Quiver Matrix Model}

A particular limit of the model we have been studying is related to
the deconstruction of the six-dimensional $(2,0)$ superconformal
theory \nima. This is also an interesting limit from the point of view
of $c=1$ theory. If we consider the uncompactified limit $k\rightarrow
\infty$ which leads to $c=1$ string on an infinite real line, then we
obtain the ${\widehat A}_\infty$ quiver theory.  The infinite array of
nodes of the quiver is naturally identified as a deconstruction, or
discretization, of the $c=1$ line.

Thus, if we consider the uncompactified $c=1$ theory with all the
momentum fields (including the cosmological constant operator) turned
off, {\it i.e.}\ $M_i=M$ for all nodes $i$, and then turn off the
superpotential $W(\Phi_i)\rightarrow 0$, we obtain the quiver theory
relevant for the deconstruction of the $(2,0)$ SCFT in $d=6$ \nima.
The choice of setting $W\rightarrow 0$ can be viewed as a
regularization of the $(2,0)$ theory (freezing to particular points on
the scalar moduli space, similarly to what was done in \cv\ in getting
$\cN=2$ information from the $\cN=1$ deformation). The most natural
superpotential for the $c=1$ at infinite radius is the quadratic
potential $W(\F)=\F^2$ which freezes $\Phi$ to zero. But we can also
deform away from this point by considering a more general $W(\F)$,
alternating in sign from one node to another, to freeze to an
arbitrary point in the $\Phi$ moduli space.  We can also deform the
$(2,0)$ theory by having varying ranks on each node of the
$A_{\infty}$ quiver. This gets mapped to the correlations of the
momentum states of $c=1$ theory on an infinite line.  Given the vast
literature on correlations of the decompactified $c=1$ string theory,
this should lead to new insights into $(2,0)$ little strings, which is
worth further study.

Let us explain further why obtaining the deconstruction theory of
$(2,0)$ superconformal theory is not unexpected. As was shown in
\kw\ the theory of D3 branes near a conifold
singularity is the same as the theory of D3 brane near an $A_1$
singularity where the $A_1$ geometry is fibered over the plane (giving
rise to the superpotential $\Tr\,\Phi^2$).  Orbifolding this by $\Z_k$
can be viewed as orbifolding the $A_1$ singularity by $\Z_k$ which is
the same as modding $\C^2$ by $\Z_{2k}$ producing an $A_{2k-1}$
singularity. Thus the $D3$ branes in this geometry are the same as
what is used in the deconstruction of $(2,0)$ SCFT. Moreover, after
deforming by the superpotential term, is the same as the theory we are
studying which is equivalent to $c=1$ at $k$ times the self-dual
radius.

Note also that the deconstruction of the $(2,0)$ theory has
effectively allowed us to describe F-terms of this theory in terms of
a matrix model which is dual to the $c=1$ (which in turn is related to
2d black hole geometry).  This throat region description is
reminiscent of holography in this context studied in
\minsei.  Here we are encountering this
directly from the gauge theory, which is equivalent in this context to
a matrix model.

\newsec{Higher-Dimensional Field Theories}

In this section we want to point out an obvious generalization to
higher-dimensional field theories of the connection between
four-dimensional $\cN=1$ gauge theories and matrix models.

\subsec{10D Super-Yang-Mills and Holomorphic Chern-Simons}

Let us start with a characteristic example. Consider ten-dimensional
super Yang-Mills theory on a space-time of the form $\R^4 \times Y$,
with $Y$ a Calabi-Yau three-fold. We want to keep all the Kaluza-Klein
modes on the internal manifold $Y$, also the massive ones, and thus
consider this as a four-dimensional supersymmetric gauge theory with
an infinite number of fields. This field theory can be regarded as an
effective field theory, with some unspecified UV completion, or we can
discretize the model by deconstructing the extra dimensions. As such
the action can be written in terms of a $d=4$ $\cN=1$ superspace
notation, using $4+6$ bosonic coordinates $(x^\m,y^a)$ and the usual
4d spinor coordinates $(\th^\a,\overline{\th}{}^{\dot\a})$.

Written in this way the action will contain D-terms, that are
integrals over $d^4\th$, and F-term, that are integrals over $d^2\th$.
More precisely, for the $d=10$ SYM theory the action takes the
following form \refs{\mss,\agw}.  Choose complex coordinates
$(y^1,y^2,y^3)$ for the internal space $Y$.  Similarly write the
internal components of the Yang-Mills gauge fields in terms of a
holomorphic connection $A_i(x;y)$ ($i=1,2,3$) and a conjugated
anti-holomorphic connection $A_{\ibar}(x;y)$. Let $\cW_\a(x;y)$ be the
spinor field strength of the four-dimensional gauge connection
$V(x;y)$. The fields $\cW_\a$ and $A_i$ should be considered as 
infinite sets of four-dimensional chiral superfields, parametrized by
the internal coordinate $y \in Y$.

The F-term of the action then takes the form \refs{\mss,\agw}
$$
\int d^4x d^2\th\ W_{\rm tree}(\cW_\a,A_i),
$$
where the tree-level superpotential is given by
\eqn\tree{
W_{\rm tree} = \int_Y d^6y  \Tr\left[
\cW_\a^2 + \epsilon^{ijk}\left(A_i \d_j A_k +
{2\over 3} A_i A_j A_k\right)\right].
}
The second term is just the holomorphic Chern-Simons action on the
manifold $Y$. Note that the variation of the Chern-Simons term is the
(holomorphic) curvature $F_{ij}$, so that after integrating out the
auxiliary fields the action contains the term $|W'|^2=|F_{ij}|^2$
which produces exactly the internal part of the Yang-Mills term.

There is of course also a D-term, that is in this case given by
\refs{\mss,\agw}
$$
\int d^4x d^4\th \int_Y d^6y\ \Tr\left(\Dbar_\ibar e^{-V} D_i e^V -
\dbar_i e^{-V} \d_i e^V \right).
$$
Here we use the notation
$$
D_i = \d_i + A_i,\qquad \d_i = \d/\d y^i.
$$

We can now apply the philosophy of \dviii\ to this model, considered
as a 4d $\cN=1$ theory, and compute the effective superpotential in
terms of an auxiliary internal field theory given by the action
\tree. So the matrix model of \dviii\ is now replaced by the
six-dimensional holomorphic Chern-Simons gauge theory. This is of course
just a direct field theory derivation of the result of \bcov\ that the
four-dimensional effective action of the type IIB open string is
computed in terms of the B-model topological open string on the
internal manifold. In this case the open string theory is given by a
collection of $N$ D9 branes wrapped over the Calabi-Yau $Y$.

Perhaps we should spell out this connection a bit more precisely in
this case. Let $M$ be the rank of the holomorphic Chern-Simons gauge
theory. (Again, just as in the case in 4 dimensions, we should be
careful not to confuse the rank of the gauge theory $N$ and the rank of
the auxiliary topological theory $M$.) Suppose that the manifold $Y$ is
such that there are $n$ isolated critical points of the
superpotential, {\it i.e.}\ $n$ inequivalent holomorphic connections
without moduli. Let $M_1,\ldots, M_n$ be the rank of these
connections.  Then we have a symmetry breaking pattern
$$
U(M) \to U(M_1) \times \cdots \times U(M_n)
$$
and can consider the 't Hooft limit $M_i \to \infty$. This will then
lead in the usual way to $n$ gaugino condensates $S_i=g_s M_i$.  The
4d effective superpotential is now again given by
\eqn\www{
W_{\rm eff} = \sum_i \left(N_i{\d \cF_0 \over \d S_i} + \t_i S_i\right)
}
with $\cF_0$ the planar partition function of the holomorphic CS field
theory.  Note that if we applied this to the case of ordinary CS
theory on $S^3$ we obtain the embedding of the duality of CS with
topological string \refs{\gv,\vaug} into a direct 4d field theory
language (for another description of this field theory see \akmv).

\subsec{General Philosophy}

The example of 10d super Yang-Mills illustrates nicely the general
philosophy. Any $4+k$-dimensional supersymmetric gauge theory can be
written in terms of 4d $\cN=1$ superfields $\F_i(x,\th;y)$
parametrized by coordinates $y^a$ on an internal $k$-dimensional
manifold $Y$. Such a theory will have a tree-level superpotential that
is given by some $k$-dimensional {\it bosonic} action
$$
W_{\rm tree} = \cS[\F_i]=  \int_Y d^ky\ {\cal L}(\Phi_i)
$$
containing in general gauge fields and matter fields on $Y$. The 4d
effective {\it superpotential} is now expressed through \www\ in terms
of the effective {\it action} $\cF_0$ of the internal $k$-dimensional
bosonic gauge theory.
$$
\lim_{M \to \infty} \int {\cal D}\F\ e^{-\cS[\F]/g_s} = e^{-\cF_0/g_s^2}
$$
For classical groups and (bi)fundamental matter
this effective action is computed in terms of planar Feynman graphs of
the internal theory.

Many interesting cases can be studied in this way.  In the next
subsection we explain how 5-dimensional $\cN=1$ supersymmetric gauge
theories can be deconstructed in this way, which leads us directly to
matrix quantum mechanics.

\subsec{Deconstruction of 5D Gauge Theories and Matrix Quantum Mechanics}

Let us first consider deconstruction of a pure $\cN=1$ Yang-Mills in 5
dimensions compactified on a circle of radius $\beta$.  This
case has been recently studied in \ika .  Viewed from
the perspective of $\cN=1$ in 4 dimensions this has in addition a
chiral scalar field $\F$ given by the component of the gauge field
along the 5th direction.  (The extra (real) scalar field, that is part
of the 5-dimensional vector multiplet, complexifies $\F$.)  This scalar
gives rise to a holonomy
$$
U=\exp( \beta \Phi) \in U(N)
$$
around the 5th circle of length $\beta$.  Here $\Phi$ (whose
eigenvalues are periodic with period $2\pi i/\beta$) represents a flat
direction.

In terms of the general discussion of the previous section the internal
theory is here a one-dimensional gauge theory. Such a theory is
entirely described in terms of the holonomy $U$ modulo conjugation by
$U(N)$. So we immediately see that according to the above philosophy
5d super Yang-Mills will be related to a {\it unitary} matrix
model. If we break to $\cN=1$ in 4 dimensions by choosing some
tree-level superpotential $W(U)$, the F-terms of this
deformed theory are computed by the large $M$ limit of the
holomorphic gauged unitary matrix model
$$
Z = {1\over \vol\, U(M)} \int dU \ 
\exp\left[ -{1\over g_s}\Tr W(U) \right],
$$
with $dU$ the Haar measure on $U(M)$.

To gain insight into freezing the moduli at a particular point, we
consider deforming the theory by a superpotential of the same degree
as the rank of the $U(N)$ group and choose the $N$ eigenvalues of
$\Phi$ to fill the $N$ distinct values of the critical points. We can
recover the information about the original theory by letting the
strength of the superpotential to go to zero, as in \cv.  In particular
we consider a superpotential $W$ of the form
$$W(U)=\sum_{i=0}^N {g_i} U^i,$$
which has $N$ distinct critical points (viewing $\Phi$ as the fundamental
field).   We have
$$
{1\over \beta}{dW \over d\Phi} = P(U)=U W'(U)=\prod_{i=1}^N (U-a_i)=0.
$$

We can now study this theory in the planar limit to extract exact
information about this theory in 4 dimensions.  The planar limit of
unitary matrix models such as this one have been analyzed before
\refs{\gw,\wadia} and reanalyzed recently in \dvii.  In particular
following the analysis of \dvii\ we find that the spectral curve is
given by
\eqn\unima{y^2+y P(u)+f(u)=0,}
where $f(u)$ is a quantum correction depending on how the eigenvalues
are distributed. It is given by the matrix model expectation value
$$
f(u) = \left\langle \Tr\left[\(P(U) - P(u)\)
{U + u\over U -u}\right]\right\rangle.
$$
Using this formula one sees immediately, as in \dvii, that $f(u)$ is a
polynomial in $u$, with degree at most $N$, {\it  i.e.}
$$f(u)=\sum_{i=0}^N b_i u^i$$
for some $b_i$.  If we distribute the eigenvalues equally among the
vacua and then extremize the superpotential (with respect to the
$S_i$'s), we expect, as in \cv\ that the quantum correction $f(u)$
will simplify. More precisely, we expect that all $b_i=0$ except for
$b_0$, so that
$$
f(u) = b_0 .
$$
This remains to be shown, which should be possible using the
techniques of \cv.

This in particular would lead to the curve of the compactified 5d
theory
\eqn\gotc{y^2+y P(u)+b_0 =
y^2+y\prod_{i=1}^N (u-a_i) +b_0 =0.}
The period matrix of this Riemann surface gives the
gauge coupling constant of the $U(1)^N$ in the 4d theory.  This is
indeed the correct curve \refs{\kkv ,\kmv}\ for a particular value of
the level $k$ of the 5d Chern-Simons term, namely $k=N$.  This is
compatible with recent results of \ika\ where they argue why the
simplest deconstruction procedure (which is equivalent to our setup)
gives rise to this particular value of Chern-Simons term in 5d. However, as
is known from \ims\ the allowed
values of $k$ leading to a decoupled 5d theory are $|k|\leq N$.
Moreover, from \refs{\kkv,\kmv}\ we know that this gives rise to
curve \unima\ with the correction term
$$
f(u)=b u^{N-|k|}.
$$
This is indeed compatible with allowed quantum corrections of the
matrix model, which suggests that, as we change the CS level, we
should be minimizing a different superpotential.  This is not
unreasonable, as the CS term involves from the 4d point of view a term
$\Phi F\wedge F$. So changing the coefficient of this term will change
the superpotential to be extremized. It would be interesting to
incorporate this into the superpotential and see why extremization now
changes $f$ to a different monomial, as is expected.

We can extend these considerations to include $\cN=1$ gauge theories
coupled to hypermultiplets in 5 dimensions. In that case we obtain a
one-dimensional gauge theory with matter fields. Now we can no longer
reduce to zero-modes and we get an honest quantum mechanical
system. For example, a five-dimensional gauge theory coupled to a set
of hypermultiplets, that we write in terms of 4d chiral multiplets as
$(Q^i,P_i)$, including some superpotential $H(Q,P)$, leads to a matrix
quantum mechanics model with action (we write the internal 5th
coordinate as $t$)
$$
\int_0^\beta dt\ \Tr \left(P_i {D Q^i \over dt} + H(P,Q)+W(U)\right)
$$ 
Here the one-dimensional gauge field $\F(t)$ appears in the covariant
derivative, although by a gauge transformation this connection can be
eliminated in favor of a boundary condition for the matter fields
twisted by the holonomy $U$. We have also included the superpotential
$W(U)$ as we did before, for the scalar field $\Phi$. (Compare the
above action with the three-matrix model \kaz.)  Note that if we take
$H(P,Q)=P^2+V(Q)$ and integrate out $P$ we get the standard gauge
quantum matrix model.  The particular case where $H(P,Q)=P^2-Q^2$ is
related in a suitable double scaling limit to the $c=1$ theory \mm.
It should be compared to  \twomm\ with $X_\pm=P\pm Q$.

\subsec{Other Examples}

This approach opens up many further interesting possibilities. For
example, we already mentioned that compactifying D6 branes on a
three-manifold $M$ gives rise to an internal Chern-Simons gauge
theory. If we compactify first on a circle, and thus take $M=S^1
\times \Sigma$, the resulting 2d action on $\Sigma$ takes a ``BF'' form
$$
\cS = \int_\Sigma \Tr\(\F F\)
$$
This is also known as 2d topological Yang-Mills \witYM. We can break the
supersymmetry further down to $\cN=1$ by deforming this to
$$
\cS = \int_\Sigma \Tr\(\F F + W(\F))\
$$ 
Taking a quadratic term gives 2d Yang-Mills. Its large $N$ limit has
been solved in \gt. It would be interesting to further explore this
connection.

In fact, the most general statement is the following. Take one's
favorite $k$-dimensional $U(N)$ (bosonic) gauge theory and promote it
to an internal superpotential of a $4+k$-dimensional theory by
interpretating all fields as chiral superfields and coupling them to
four-dimensional gauge fields. Choose some appropriate D-terms to
complete the theory. Then the planar diagrams of the chosen gauge
theory will compute all F-terms in the four-dimensional effective
action. In these more general internal bosonic gauge theories one does
not get a theory which can be viewed as a KK reduction of a standard
higher-dimensional gauge theory (for example for 4d bosonic YM as internal
theory one gets $(D_i F_{ij})^2$ as the relevant piece of the action
for the extra 4 dimensions).  Nevertheless they give rise to a vast
collection of potentially interesting $\cN=1$ supersymmetric
gauge theories in 4 dimensions.

\newsec{Recovering The Full String Theory}

Although significant progress has been made in understanding the
structure of F-terms in $\cN=1$ supersymmetric gauge theories, one can
ask what can be said about the non-holomorphic information
(D-terms). In the most general case there is little hope that these
terms are under control. But a special role should be played by
superconformal fixed points. In these cases one can expect that
F-terms are enough to completely specify the theory. If that is true,
this would be the strongest possible sense in which $\cN=1$ systems
are solvable.

This is a well-established fact for $\cN=2$ superconformal field
theories in two dimensions. In that case the only marginal
deformations are given by variations of F-terms. The proof is
straightforward --- D-term variations are of the form
$Q^2\overline{Q}^2 \F$, with $\F$ an operator of conformal dimension
zero and thus equal to the identity. The only possible deformations
are therefore F-terms. It would be interesting to find a proof for the
corresponding statement in the four-dimensional case.  To the best of
our knowledge there is no counterexample to this claim.

In the present case it is clear that the holomorphic data uniquely
specify the corresponding string theory. Given the complex structure
and fluxes the (generalized) Calabi-Yau geometry is fixed by the
world-sheet beta-functions. Furthermore this data is entirely captured
by the matrix integral. In that sense we can say that the matrix model
is a very efficient, though implicit, way to encode the full string
theory.

\bigskip
\centerline{\bf Acknowledgements}

We would like to thank M. Aganagic, N. Arkani-Hamed, J. de Boer,
S. Gukov, K. Intriligator, A. Iqbal, V. Kazakov, I. Kostov, H. Ooguri,
L. Motl, A. Sinkovics, K. Skenderis, and N. Warner for valuable
discussions. Furthermore C.V. wishes to thank the hospitality of the
Physics Department at Caltech, where he is a Moore
Scholar.

The research of R.D.~is partly supported by FOM and the CMPA grant
of the University of Amsterdam, C.V.~is partly supported by NSF
grants PHY-9802709 and DMS-0074329.

\listrefs

\end